\documentclass[12pt,reqno]{amsart}%
\usepackage{graphicx}
\usepackage{amscd}
\usepackage{amsmath}
\usepackage{epsfig}
\usepackage{amsfonts}
\usepackage{amssymb}%
\providecommand{\U}[1]{\protect\rule{.1in}{.1in}}
\providecommand{\U}[1]{\protect\rule{.1in}{.1in}}
\providecommand{\U}[1]{\protect\rule{.1in}{.1in}}
\allowdisplaybreaks

\theoremstyle{plain}

\numberwithin{equation}{section}

\begin{document}
\title[Forced Harmonic Oscillator]{The Cauchy Problem for a Forced Harmonic Oscillator }
\author{Raquel M.~Lopez}
\address{Department of Mathematics and Statistics, Arizona State University, Tempe, AZ
85287--1804, U.S.A.}
\email{rlopez14@asu.edu}
\author{Sergei K. Suslov}
\address{Department of Mathematics and Statistics, Arizona State University, Tempe, AZ
85287--1804, U.S.A.}
\email{sks@asu.edu}
\urladdr{http://hahn.la.asu.edu/\symbol{126}suslov/index.html}
\date{\today}

\begin{abstract}
We construct an explicit solution of the Cauchy initial value problem for the
one-dimensional Schr\"{o}dinger equation with a time-dependent Hamiltonian
operator for the forced harmonic oscillator. The corresponding Green function
(propagator) is derived with the help of the generalized Fourier transform and
a relation with representations of the Heisenberg--Weyl group $N\left(
3\right)  $ in a certain special case first, and then is extended to the
general case. A three parameter extension of the classical Fourier integral is
discussed as a by-product. Motion of a particle with a spin in uniform
perpendicular magnetic and electric fields is considered as an application; a
transition amplitude between Landau levels is evaluated in terms of Charlier
polynomials. In addition, we also solve an initial value problem to a similar
diffusion-type equation.

\end{abstract}
\maketitle

\section{Introduction}

The time-dependent Schr\"{o}dinger equation for the one-dimensional harmonic
oscillator has the form%
\begin{equation}
i\hslash\frac{\partial\psi}{\partial t}=H\psi, \label{int1}%
\end{equation}
where the Hamiltonian is
\begin{equation}
H=\frac{\hslash\omega}{2}\left(  -\frac{\partial^{2}}{\partial x^{2}}%
+x^{2}\right)  =\frac{\hslash\omega}{2}\left(  aa^{\dagger}+a^{\dagger
}a\right)  . \label{int2}%
\end{equation}
Here $a^{\dagger}$ and $a$ are the creation and annihilation operators,
respectively, given by%
\begin{equation}
a^{\dagger}=\frac{1}{\sqrt{2}}\left(  x-\frac{\partial}{\partial x}\right)
,\qquad a=\frac{1}{\sqrt{2}}\left(  x+\frac{\partial}{\partial x}\right)  ;
\label{int3}%
\end{equation}
see \cite{Flu} for another definition. They satisfy the familiar commutation
relation%
\begin{equation}
\left[  a,\ a^{\dagger}\right]  =aa^{\dagger}-a^{\dagger}a=1. \label{int4}%
\end{equation}
A natural modification of the Hamiltonian operator (\ref{int2}) is as follows%
\begin{equation}
H\rightarrow H\left(  t\right)  =\frac{\hslash\omega}{2}\left(  aa^{\dagger
}+a^{\dagger}a\right)  +\hslash\left(  \delta\left(  t\right)  a+\delta^{\ast
}\left(  t\right)  a^{\dagger}\right)  , \label{int5}%
\end{equation}
where $\delta\left(  t\right)  $ is a complex valued function of time $t$ and
the symbol $\ast$ denotes complex conjugation. This operator is Hermitian,
namely, $H^{\dagger}\left(  t\right)  =H\left(  t\right)  .$ It corresponds to
the case of the forced harmonic oscillator which is of interest in many
advanced problems. Examples include polyatomic molecules in varying external
fields, crystals through which an electron is passing and exciting the
oscillator modes, and other interactions of the modes with external fields. It
has particular applications in quantum electrodynamics because the
electromagnetic field can be represented as a set of forced harmonic
oscillators \cite{Bo:Shi}, \cite{Fey:Hib}, \cite{Gottf:T-MY}, \cite{Merz},
\cite{Mes} and \cite{Schiff}. Extensively used propagator techniques were
originally introduced by Richard Feynman in \cite{FeynmanPhD}, \cite{Feynman},
\cite{Feynman49a} and \cite{Feynman49b}.\smallskip

On this note we construct an exact solution of the time-dependent
Schr\"{o}dinger equation%
\begin{equation}
i\hslash\frac{\partial\psi}{\partial t}=H\left(  t\right)  \psi\label{int6}%
\end{equation}
with the Hamiltonian of the form (\ref{int5}), subject to the initial
condition%
\begin{equation}
\left.  \psi\left(  x,t\right)  \right\vert _{t=0}=\psi_{0}\left(  x\right)  ,
\label{int7}%
\end{equation}
where $\psi_{0}\left(  x\right)  $ is an arbitrary square integrable complex
valued function from $\mathcal{L}^{2}\left(  -\infty,\infty\right)  .$ We
shall start with a particular choice of the time-dependent function
$\delta\left(  t\right)  $ given by (\ref{separ6}) below, which is later
extended to the general case. The explicit form of equation (\ref{int6}) is
given by (\ref{solu1}) and (\ref{fho1}) below, and an extension to similar
diffusion-type equations is also discussed.\smallskip

This paper is organized as follows. In section~2 we remind the reader about
the textbook solution of the stationary Schr\"{o}dinger equation for the
one-dimentional simple harmonic oscillator. In section~3 we consider the
eigenfunction expansion for the time-dependent Schr\"{o}dinger equation
(\ref{int6}) and find its particular solutions in terms of the Charlier
polynomials for certain forced harmonic oscillator. The series solution of the
corresponding initial value problem is obtained in section~4. It is further
transformed into an integral form in section~7 after discussing two relevant
technical tools, namely, the representations of the Heisenberg--Weyl group
$N\left(  3\right)  $ and the generalized Fourier transform in sections~5 and
6, respectively. An important special case of the Cauchy initial value problem
for the simple harmonic oscillator is outlined in section~8 and a three
parameter generalization of the Fourier transform is introduced in section~9
as a by-product. In sections~10 and 11 we solve the initial value problem for
the general forced harmonic oscillator in terms of the corresponding Green
function (or Feynman's propagator) and the eigenfunction expansion,
respectively, by a different method that uses all technical tools developed
before in the special case. An extension to the case of time-dependent
frequency is given in section~12. Then in section~13, we outline important
special and limiting cases of the Feynman propagators. Finally in section~14,
the motion of a charged particle with a spin in uniform magnetic and electric
fields that are perpendicular to each other is considered as an application;
we evaluate a transition amplitude between Landau levels under the influence
of the perpendicular electric field in terms of Charlier polynomials and find
the corresponding propagator in three dimensions. Solutions of similar
diffution-type equations are discussed in section~15.\smallskip

The Cauchy initial value problem for a forced harmonic oscillator was
originally considered by Feynman in his path integrals approach to the
nonrelativistic quantum mechanics \cite{FeynmanPhD}, \cite{Feynman}, and
\cite{Fey:Hib}. Since then this problem and its special and limiting cases
were discussed by many authors \cite{Beauregard}, \cite{Gottf:T-MY},
\cite{Holstein}, \cite{Maslov:Fedoriuk}, \cite{Merz}, \cite{Thomber:Taylor}
the simple harmonic oscillator; \cite{Arrighini:Durante}, \cite{Brown:Zhang},
\cite{Holstein97}, \cite{Nardone}, \cite{Robinett} the particle in a constant
external field; see also references therein. It is worth noting that an exact
solution of the $n$-dimensional time-dependent Schr\"{o}dinger equation for
certain modified oscillator is found in \cite{Me:Co:Su}. These simple exactly
solvable models may be of interest in a general treatment of the non-linear
time-dependent Schr\"{o}dinger equation (see \cite{Howland}, \cite{Jafaev},
\cite{Naibo:Stef}, \cite{Rod:Schlag}, \cite{Schlag}, \cite{Yajima} and
references therein). They also provide explicit solutions which can be useful
for testing numerical methods of solving the time-dependent Schr\"{o}dinger equation.

\section{The Simple Harmonic Oscillator in One Dimension}

The time-dependent Hamiltonian operator (\ref{int5}) has the following
structure%
\begin{equation}
H\left(  t\right)  =H_{0}+H_{1}\left(  t\right)  , \label{harm1}%
\end{equation}
where%
\begin{equation}
H_{0}=\frac{\hslash\omega}{2}\left(  aa^{\dagger}+a^{\dagger}a\right)
\label{harm2}%
\end{equation}
is the Hamiltonian of the harmonic oscillator and%
\begin{equation}
H_{1}\left(  t\right)  =\hslash\left(  \delta\left(  t\right)  a+\delta^{\ast
}\left(  t\right)  a^{\dagger}\right)  \label{harm3}%
\end{equation}
is the time-dependent \textquotedblleft perturbation\textquotedblright, which
corresponds to an external time-dependent force that does not depend on the
coordinate $x$ (dipole interaction) and a similar velocity-dependent term (see
\cite{Fey:Hib}, \cite{Gottf:T-MY} and \cite{Merz} for more details).\smallskip

The solution of the stationary Schr\"{o}dinger equation for the
one-dimensional harmonic oscillator%
\begin{equation}
H_{0}\Psi=E\Psi,\qquad H_{0}=\frac{\hslash\omega}{2}\left(  -\frac
{\partial^{2}}{\partial x^{2}}+x^{2}\right)  \label{harm4}%
\end{equation}
is a standard textbook problem in quantum mechanics (see \cite{Dav},
\cite{Flu}, \cite{La:Lif}, \cite{Merz}, \cite{Mes}, \cite{Ni:Uv},
\cite{Schiff}, and \cite{Wi} for example). The orthonormal wave functions are
given by%
\begin{equation}
\Psi=\Psi_{n}\left(  x\right)  =\frac{1}{\sqrt{2^{n}n!\sqrt{\pi}}}%
\ e^{-x^{2}/2}H_{n}\left(  x\right)  \label{harm5}%
\end{equation}
with%
\begin{equation}
\int_{-\infty}^{\infty}\Psi_{n}^{\ast}\left(  x\right)  \Psi_{m}\left(
x\right)  \ dx=\delta_{nm}=\left\{
\begin{array}
[c]{c}%
1,\qquad n=m,\medskip\\
0,\qquad n\neq m,
\end{array}
\right.  \label{harm6}%
\end{equation}
where $H_{n}\left(  x\right)  $ are the Hermite polynomials, a family of the
(very) classical orthogonal polynomials (see \cite{An:As}, \cite{An:As:Ro},
\cite{Askey}, \cite{Chihara}, \cite{Erd}, \cite{Ni:Su:Uv}, \cite{Ni:Uv},
\cite{Rainville}, and \cite{Sze}). The corresponding oscillator discrete
energy levels are%
\begin{equation}
E=E_{n}=\hslash\omega\left(  n+\frac{1}{2}\right)  \qquad\left(
n=0,1,2,...\ \right)  . \label{harm7}%
\end{equation}
The actions of the creation and annihilation operators (\ref{int3}) on the
oscillator wave functions (\ref{harm5}) are given by%
\begin{equation}
a\ \Psi_{n}=\sqrt{n}\ \Psi_{n-1},\qquad a^{\dagger}\ \Psi_{n}=\sqrt{n+1}%
\ \Psi_{n+1}. \label{harm8}%
\end{equation}
These \textquotedblleft ladder\textquotedblright\ equations follow from the
differentiation formulas%
\begin{equation}
\frac{d}{dx}H_{n}\left(  x\right)  =2nH_{n-1}\left(  x\right)  =2xH_{n}\left(
x\right)  -H_{n+1}\left(  x\right)  , \label{harm9}%
\end{equation}
which are valid for the Hermite polynomials.

\section{Eigenfunction Expansion for the Time-Dependent Schr\"{o}dinger
Equation}

In spirit of Dirac's time-dependent perturbation theory in quantum mechanics
(see \cite{Dav}, \cite{Flu}, \cite{La:Lif}, \cite{Mes}, and \cite{Schiff}) we
are looking for a solution to the initial value problem in (\ref{int6}%
)--(\ref{int7}) as an infinite series%
\begin{equation}
\psi=\psi\left(  x,t\right)  =\sum_{n=0}^{\infty}c_{n}\left(  t\right)
\ \Psi_{n}\left(  x\right)  , \label{separ1}%
\end{equation}
where $\Psi_{n}\left(  x\right)  $ are the oscillator wave functions
(\ref{harm5}) which depend only on the space coordinate $x$ and $c_{n}\left(
t\right)  $ are the yet unknown time-dependent coefficients. Substituting this
form of solution into the Schr\"{o}dinger equation (\ref{int6}) with the help
of the orthogonality property (\ref{harm6}) and the \textquotedblleft
ladder\textquotedblright\ relations (\ref{harm8}), we obtain the following
linear infinite system%
\begin{equation}
i\frac{dc_{n}\left(  t\right)  }{dt}=\omega\left(  n+\frac{1}{2}\right)
c_{n}\left(  t\right)  +\delta\left(  t\right)  \sqrt{n+1}\ c_{n+1}\left(
t\right)  +\delta^{\ast}\left(  t\right)  \sqrt{n}\ c_{n-1}\left(  t\right)
\quad\left(  n=0,1,2,...\ \right)  \label{separ2}%
\end{equation}
of the first order ordinary differential equations with $c_{-1}\left(
t\right)  \equiv0.$ The initial conditions are%
\begin{equation}
c_{n}\left(  0\right)  =\int_{-\infty}^{\infty}\Psi_{n}^{\ast}\left(
x\right)  \psi_{0}\left(  x\right)  \ dx \label{separ3}%
\end{equation}
due to the initial data (\ref{int7}) and the orthogonality property
(\ref{harm6}).\smallskip

Now we specify the exact form of the function $\delta\left(  t\right)  $ in
order to find a particular solution of the system (\ref{separ2}) in terms of
the so-called Charlier polynomials that belong to the classical orthogonal
polynomials of a discrete variable (see \cite{Charlier}, \cite{Chihara},
\cite{Erd}, \cite{Ni:Su:Uv}, and \cite{Ni:Uv}). One can easily verify that the
following Ansatz%
\begin{equation}
c_{n}\left(  t\right)  =\left(  -1\right)  ^{n}\frac{\mu^{n/2}}{\sqrt{n!}%
}\ e^{-i\left(  \omega\left(  n+1/2\right)  -\left(  n+\mu\right)  \right)
t}\ \left(  \left.  e^{-i\lambda t}p_{n}\left(  \lambda\right)  \right.
\right)  \label{separ4}%
\end{equation}
gives the three term recurrence relation%
\begin{equation}
\lambda p_{n}\left(  \lambda\right)  =-\mu p_{n+1}\left(  \lambda\right)
+\left(  n+\mu\right)  p_{n}\left(  \lambda\right)  -np_{n-1}\left(
\lambda\right)  \label{separ5}%
\end{equation}
for the Charlier polynomials $p_{n}\left(  \lambda\right)  =c_{n}^{\mu}\left(
\lambda\right)  $ (see \cite{Ni:Su:Uv} and \cite{Ni:Uv} for example), when we
choose%
\begin{equation}
\delta\left(  t\right)  =\sqrt{\mu}\ e^{i\left(  \omega-1\right)  t}%
,\qquad\delta^{\ast}\left(  t\right)  =\sqrt{\mu}\ e^{-i\left(  \omega
-1\right)  t} \label{separ6}%
\end{equation}
with the real parameter $\mu$ such that $0<\mu<1.$ Thus with $p_{n}\left(
\lambda\right)  =c_{n}^{\mu}\left(  \lambda\right)  ,$ equation (\ref{separ4})
yields a particular solution of the system (\ref{separ2}) for any value of the
spectral parameter $\lambda.$\smallskip

By the superposition principle the solution to this linear system of ordinary
differential equations, which satisfies the initial condition (\ref{separ3}),
can be constructed as a linear combination%
\begin{equation}
c_{n}\left(  t\right)  =\sum_{m=0}^{\infty}c_{nm}\left(  t\right)
\ c_{m}\left(  0\right)  , \label{separ7}%
\end{equation}
where $c_{nm}\left(  t\right)  $ is a \textquotedblleft
Green\textquotedblright\ function, or a particular solution that satisfies the
simplest initial conditions%
\begin{equation}
c_{nm}\left(  0\right)  =\delta_{nm}. \label{separ8}%
\end{equation}
In the next section we will obtain the function $c_{nm}\left(  t\right)  $ in
terms of the Charlier polynomials; see equation (\ref{solu8}) below. In
section~5 we establish a relation with the representations of the
Heisenberg--Weyl group $N\left(  3\right)  ;$ see equation (\ref{heis7}). A
generalization to an arbitrary function $\delta\left(  t\right)  $ will be
given later.

\section{Solution of the Cauchy Problem}

We can now construct the exact solution to the original Cauchy problem in
(\ref{int6})--(\ref{int7}) for the time-dependent Schr\"{o}dinger equation
with the Hamiltonian of the form (\ref{harm1})--(\ref{harm3}) and
(\ref{separ6}). More explicitly, we will solve the following partial
differential equation%
\begin{equation}
i\frac{\partial\psi}{\partial t}=\frac{\omega}{2}\left(  -\frac{\partial
^{2}\psi}{\partial x^{2}}+x^{2}\psi\right)  +\sqrt{2\mu}\left(  \left(
\cos\left(  \omega-1\right)  t\right)  \ x\psi+i\left(  \sin\left(
\omega-1\right)  t\right)  \ \frac{\partial\psi}{\partial x}\right)
\label{solu1}%
\end{equation}
subject to the initial condition%
\begin{equation}
\left.  \psi\left(  x,t\right)  \right\vert _{t=0}=\psi_{0}\left(  x\right)
\qquad\left(  -\infty<x<\infty\right)  . \label{solu2}%
\end{equation}
By (\ref{separ1}), (\ref{separ3}), and (\ref{separ7}) our solution has the
form%
\begin{equation}
\psi\left(  x,t\right)  =\sum_{n=0}^{\infty}\Psi_{n}\left(  x\right)
\sum_{m=0}^{\infty}c_{nm}\left(  t\right)  \ \int_{-\infty}^{\infty}\Psi
_{m}\left(  y\right)  \psi_{0}\left(  y\right)  \ dy, \label{solu3}%
\end{equation}
where%
\begin{align}
c_{nm}\left(  t\right)   &  =\left(  -\mu^{1/2}\right)  ^{n-m}\ \sqrt
{\frac{m!}{n!}}\ e^{-i\left(  \omega\left(  n+1/2\right)  -\left(
n+\mu\right)  \right)  t}\label{solu4}\\
&  \quad\ \times\frac{\mu^{m}}{m!}\ \sum_{k=0}^{\infty}e^{-ikt}c_{n}^{\mu
}\left(  k\right)  \ c_{m}^{\mu}\left(  k\right)  \ e^{-\mu}\frac{\mu^{k}}%
{k!}\nonumber\\
&  =\left(  -1\right)  ^{n-m}\ e^{-\mu}\ \frac{\mu^{\left(  n+m\right)  /2}%
}{\sqrt{n!m!}}\ e^{-i\left(  \omega\left(  n+1/2\right)  -\left(
n+\mu\right)  \right)  t}\nonumber\\
&  \quad\ \times\sum_{k=0}^{\infty}c_{n}^{\mu}\left(  k\right)  \ c_{m}^{\mu
}\left(  k\right)  \ \frac{\left(  \mu e^{-it}\right)  ^{k}}{k!},\nonumber
\end{align}
in view of the superposition principle and the orthogonality property%
\begin{equation}
\sum_{k=0}^{\infty}c_{n}^{\mu}\left(  k\right)  c_{m}^{\mu}\left(  k\right)
\ e^{-\mu}\frac{\mu^{k}}{k!}=\frac{m!}{\mu^{m}}\ \delta_{nm}\qquad\left(
0<\mu<1\right)  \label{solu5}%
\end{equation}
of the Charlier polynomials (see \cite{Ni:Su:Uv} for example).\smallskip

The right-hand side of (\ref{solu4}) can be transformed into a single sum with
the help of the following generating relation for the Charlier polynomials%
\begin{align}
&  \sum_{k=0}^{\infty}\frac{\left(  \mu_{1}\mu_{2}s\right)  ^{k}}{k!}%
\ c_{n}^{\mu_{1}}\left(  k\right)  \ c_{m}^{\mu_{2}}\left(  k\right)
=e^{\mu_{1}\mu_{2}s}\left(  1-\mu_{1}s\right)  ^{m}\left(  1-\mu_{2}s\right)
^{n}\label{solu6}\\
&  \qquad\qquad\qquad\times\ _{2}F_{0}\left(  -n,\ -m;\ \frac{s}{\left(
1-\mu_{1}s\right)  \left(  1-\mu_{2}s\right)  }\right)  .\nonumber
\end{align}
(See \cite{Meixner34}, \cite{Meixner38}, \cite{Meixner42}, and \cite{Gasper73}
for more information and \cite{Ba} for the definition of the generalized
hypergeometric series.) Choosing $\mu_{1}=\mu_{2}=\mu$ and $s=e^{-it}/\mu$ we
obtain%
\begin{align}
c_{nm}\left(  t\right)   &  =\frac{\left(  -i\right)  ^{n+m}}{\sqrt
{2^{n+m}n!m!}}\ e^{-i\left(  \left(  \omega-1\right)  n+\left(  n+m\right)
/2\right)  t}\ e^{-2\mu\sin^{2}\left(  t/2\right)  }\ e^{-i\left(  \mu\sin
t+\left(  \omega/2-\mu\right)  t\right)  }\label{solu7}\\
&  \quad\times\left(  2\sqrt{2\mu}\sin\left(  t/2\right)  \right)
^{n+m}\ _{2}F_{0}\left(  -n,\ -m;\ -\frac{1}{4\mu\sin^{2}\left(  t/2\right)
}\right)  .\nonumber
\end{align}
The hypergeometric series representation for the Charlier polynomials is%
\begin{equation}
c_{n}^{\mu}\left(  x\right)  =\ _{2}F_{0}\left(  -n,\ -x;\ -\frac{1}{\mu
}\right)  \label{charlier1}%
\end{equation}
(see \cite{Ni:Su:Uv} for example). Thus%
\begin{align}
&  c_{nm}\left(  t\right)  =e^{-i\left(  \mu\sin t+\left(  \omega
/2-\mu\right)  t\right)  }\ e^{-i\left(  \left(  \omega-1\right)  n+\left(
n+m\right)  /2\right)  t}\label{solu8}\\
&  \qquad\qquad\times\frac{\left(  -i\right)  ^{n+m}}{\sqrt{2^{n+m}n!m!}%
}\ e^{-\beta^{2}/4}\ \beta^{n+m}\ c_{n}^{\beta^{2}/2}\left(  m\right)
\nonumber
\end{align}
with $\beta=\beta\left(  t\right)  =2\sqrt{2\mu}\sin\left(  t/2\right)  $ and,
as a result, by substitution of this expression into the series (\ref{solu3}),
we obtain the eigenfunction expansion solution to the original Cauchy problem
(\ref{solu1})--(\ref{solu2}). We shall be able to find an integral form of
this solution in section~7 after discussing representations of the
Heisenberg--Weyl group $N\left(  3\right)  $ and a generalization of the
Fourier transform in the next two sections. This complete solution of the
particular initial value problem (\ref{solu1})--(\ref{solu2}) will suggest a
correct form of the Green function (propagator) for the general forced
harmonic oscillator in sections~10 and 11.

\section{Relation with the Heisenberg--Weyl Group $N\left(  3\right)  $}

Let $N\left(  3\right)  $ be the three-dimensional group of the upper
triangular real matrices of the form%
\begin{equation}%
\begin{pmatrix}
1 & \alpha & \gamma\\
0 & 1 & \beta\\
0 & 0 & 1
\end{pmatrix}
=\left(  \alpha,\beta,\gamma\right)  . \label{heis1}%
\end{equation}
The map%
\begin{equation}
T\left(  \alpha,\beta,\gamma\right)  \Psi\left(  x\right)  =e^{i\left(
\gamma+\beta x\right)  }\ \Psi\left(  x+\alpha\right)  \label{heis2}%
\end{equation}
defines a unitary representation of the Heisenberg--Weyl group $N\left(
3\right)  $ in the space of square integrable functions $\Psi\in
\mathcal{L}^{2}\left(  -\infty,\infty\right)  $ (see \cite{Vil},
\cite{Ni:Su:Uv}, and \cite{SteinHarm} for more details).\smallskip

The set $\left\{  \Psi_{n}\left(  x\right)  \right\}  _{n=0}^{\infty}$ of the
wave functions of the harmonic oscillator (\ref{harm5}) forms a complete
orthonormal system in $\mathcal{L}^{2}\left(  -\infty,\infty\right)  .$ The
matrix elements of the representation (\ref{heis2}) with respect to this basis
are related to the Charlier polynomials as follows%
\begin{equation}
T\left(  \alpha,\beta,\gamma\right)  \Psi_{n}\left(  x\right)  =\sum
_{m=0}^{\infty}T_{mn}\left(  \alpha,\beta,\gamma\right)  \ \Psi_{m}\left(
x\right)  , \label{heis3}%
\end{equation}
where%
\begin{align}
&  T_{mn}\left(  \alpha,\beta,\gamma\right)  =\int_{-\infty}^{\infty}\Psi
_{m}^{\ast}\left(  x\right)  e^{i\left(  \gamma+\beta x\right)  }\ \Psi
_{n}\left(  x+\alpha\right)  \ dx\label{heis4}\\
&  \quad=\frac{i^{m-n}}{\sqrt{m!n!}}\ e^{i\left(  \gamma-\alpha\beta/2\right)
}\ e^{-\nu/2}\ \left(  \frac{i\alpha+\beta}{\sqrt{2}}\right)  ^{m}\left(
\frac{i\alpha-\beta}{\sqrt{2}}\right)  ^{n}\ c_{m}^{\nu}\left(  n\right)
\nonumber
\end{align}
with $\nu=\left(  \alpha^{2}+\beta^{2}\right)  /2$ \cite{Ni:Su:Uv}. A similar
integral%
\begin{equation}
\int_{-\infty}^{\infty}H_{m}\left(  x+y\right)  H_{n}\left(  x+z\right)
\ e^{-x^{2}}dx=\sqrt{\pi}\ 2^{n}m!\ z^{n-m}\ L_{m}^{n-m}\left(  -2yz\right)
\label{heis3a}%
\end{equation}
is evaluated in \cite{ErdInt} in terms of the Laguerre polynomials
$L_{m}^{\alpha}\left(  \xi\right)  ,$ whose relation with the Charlier
polynomials is%
\begin{equation}
c_{n}^{\mu}\left(  x\right)  =\left(  -\mu\right)  ^{-n}n!L_{n}^{x-n}\left(
\mu\right)  , \label{heis3aa}%
\end{equation}
see \cite{Ni:Su:Uv}. Its special case $y=z$ in the form of%
\begin{equation}
\int_{-\infty}^{\infty}H_{m}\left(  x\right)  H_{n}\left(  x\right)
\ e^{-\left(  x-y\right)  ^{2}}dx=\sqrt{\pi}\ 2^{n}m!\ y^{n-m}\ L_{m}%
^{n-m}\left(  -2y^{2}\right)  \label{heis3ab}%
\end{equation}
is of a particular interest in this paper.\smallskip

The unitary relation%
\begin{equation}
\sum_{n=0}^{\infty}T_{mn}^{\ast}\left(  \alpha,\beta,\gamma\right)
T_{m^{\prime}n}\left(  \alpha,\beta,\gamma\right)  =\delta_{mm^{\prime}}
\label{heis4a}%
\end{equation}
holds due to the orthogonality property of the Charlier polynomials
(\ref{solu5}).\smallskip

The relevant special case of these matrix elements is%
\begin{equation}
T_{mn}\left(  0,\beta,0\right)  =t_{mn}\left(  \beta\right)  =\frac{i^{m+n}%
}{\sqrt{2^{m+n}m!n!}}\ e^{-\beta^{2}/4}\ \beta^{m+n}\ c_{m}^{\beta^{2}%
/2}\left(  n\right)  , \label{heis5}%
\end{equation}
which explicitly acts on the oscillator wave functions as follows%
\begin{equation}
e^{i\beta x}\Psi_{n}\left(  x\right)  =\sum_{m=0}^{\infty}t_{mn}\left(
\beta\right)  \ \Psi_{m}\left(  x\right)  . \label{heis6}%
\end{equation}
Relations (\ref{solu6}), (\ref{charlier1}) and (\ref{heis5}) imply%
\begin{align}
&  \sum_{k=0}^{\infty}t_{mk}\left(  \beta_{1}\right)  t_{nk}\left(  \beta
_{2}\right)  \ s^{k}=\frac{i^{m+n}}{\sqrt{2^{m+n}m!n!}}\ e^{-\left(  \beta
_{1}^{2}+\beta_{2}^{2}+2\beta_{1}\beta_{2}s\right)  /4}\label{heis6a}\\
&  \qquad\qquad\qquad\qquad\qquad\quad\times\left(  \beta_{1}+\beta
_{2}s\right)  ^{m}\left(  \beta_{2}+\beta_{1}s\right)  ^{n}\ c_{m}^{\lambda
}\left(  n\right) \nonumber
\end{align}
with $\lambda=\left(  \beta_{1}^{2}+\beta_{2}^{2}+\beta_{1}\beta_{2}\left(
s+s^{-1}\right)  \right)  /2,$ which is an extension of the addition formula%
\begin{equation}
\sum_{k=0}^{\infty}t_{mk}\left(  \beta_{1}\right)  t_{nk}\left(  \beta
_{2}\right)  =t_{mn}\left(  \beta_{1}+\beta_{2}\right)  \label{heis6b}%
\end{equation}
for the matrix elements.\smallskip

In order to obtain functions $c_{nm}\left(  t\right)  $ in terms of the matrix
elements $t_{mn}\left(  \beta\right)  $ of the representations of the
Heisenberg--Weyl group, we compare (\ref{solu8}) and (\ref{heis5}). The result
is%
\begin{equation}
c_{nm}\left(  t\right)  =\left(  -1\right)  ^{n+m}e^{-i\left(  \mu\sin
t+\left(  \omega/2-\mu\right)  t\right)  }\ e^{-i\left(  \left(
\omega-1\right)  n+\left(  n+m\right)  /2\right)  t}\ t_{mn}\left(
\beta\right)  , \label{heis7}%
\end{equation}
where $\beta=2\sqrt{2\mu}\sin\left(  t/2\right)  .$ Our solution (\ref{solu3})
takes the form%
\begin{align}
\psi\left(  x,t\right)   &  =\sum_{n=0}^{\infty}\Psi_{n}\left(  x\right)
\sum_{m=0}^{\infty}c_{nm}\left(  t\right)  \ \int_{-\infty}^{\infty}\Psi
_{m}\left(  y\right)  \psi_{0}\left(  y\right)  \ dy\label{heis8}\\
&  =e^{-i\left(  \mu\sin t+\left(  \omega/2-\mu\right)  t\right)  }\sum
_{n=0}^{\infty}\left(  -1\right)  ^{n}e^{-it\left(  \omega-1/2\right)
n}\ \Psi_{n}\left(  x\right) \nonumber\\
&  \quad\times\sum_{m=0}^{\infty}t_{mn}\left(  \beta\right)  \int_{-\infty
}^{\infty}\left(  -1\right)  ^{m}e^{-imt/2}\ \Psi_{m}\left(  y\right)
\psi_{0}\left(  y\right)  \ dy.\nonumber
\end{align}
In the next section, we will discuss a generalization of the Fourier
transformation, which will allow us to transform this multiple series into a
single integral form in the section~7; see equations (\ref{green4}) and
(\ref{green8})--(\ref{green10}) below.

\section{The Generalized Fourier Transform}

The Mehler generating function, or the Poisson kernel for Hermite polynomials,
is given by%
\begin{equation}
K_{r}\left(  x,y\right)  =\sum_{n=0}^{\infty}r^{n}\ \Psi_{n}\left(  x\right)
\Psi_{n}\left(  y\right)  =\frac{1}{\sqrt{\pi\left(  1-r^{2}\right)  }}%
\ \exp\left(  \frac{4xyr-\left(  x^{2}+y^{2}\right)  \left(  1+r^{2}\right)
}{2\left(  1-r^{2}\right)  }\right)  , \label{four1}%
\end{equation}
where $\Psi_{n}\left(  z\right)  $ are the oscillator wave functions defined
by (\ref{harm5}) and $\left\vert r\right\vert \leq1,$ $r\neq\pm1$ (see
\cite{Erd}, \cite{Rainville}, \cite{Sze}, and \cite{Wi} for example). Using
the orthogonality property (\ref{harm6}) one gets%
\begin{equation}
r^{n}\Psi_{n}\left(  x\right)  =\int_{-\infty}^{\infty}K_{r}\left(
x,y\right)  \Psi_{n}\left(  y\right)  \ dy,\qquad\left\vert r\right\vert <1.
\label{four1a}%
\end{equation}
Thus the wave functions $\Psi_{n}$ are also eigenfunctions of an integral
operator corresponding to the eigenvalues $r^{n}.$\smallskip

We denote%
\begin{equation}
\mathcal{K}_{\tau}\left(  x,y\right)  =K_{e^{i\tau}}\left(  x,y\right)
=\frac{e^{i\left(  \pi/2-\tau\right)  /2}}{\sqrt{2\pi\sin\tau}}\ \exp\left(
i\frac{2xy-\left(  x^{2}+y^{2}\right)  \cos\tau}{2\sin\tau}\right)
\label{four2}%
\end{equation}
with $0<\tau<\pi$ and use the fact that the oscillator wave functions are the
eigenfunctions of the generalized Fourier transform%
\begin{equation}
e^{in\tau}\Psi_{n}\left(  x\right)  =\int_{-\infty}^{\infty}\mathcal{K}_{\tau
}\left(  x,y\right)  \Psi_{n}\left(  y\right)  \ dy \label{four3}%
\end{equation}
corresponding to the eigenvalues $e^{in\tau}.$ (See \cite{As:Ra:Su},
\cite{Ra:Su} and \cite{SusAlg} for more details on the generalized Fourier
transform, its inversion formula and their extensions. It is worth noting that
the classical Fourier transform corresponds to the particular value $\tau
=\pi/2$ \cite{Wi}. Its three parameter extension will be discussed in section~9.)

\section{An Integral Form of Solution}

Now let us transform the series (\ref{heis8}) into a single integral form.
With the help of the inversion formula for the generalized Fourier transform
(see (\ref{four3}) with $\tau\rightarrow-\tau$) and the symmetry property%
\[
H_{n}\left(  -x\right)  =\left(  -1\right)  ^{n}\ H_{n}\left(  x\right)
\]
we get%
\begin{equation}
\left(  -1\right)  ^{m}e^{-imt/2}\ \Psi_{m}\left(  y\right)  =\int_{-\infty
}^{\infty}\mathcal{K}_{-t/2}\left(  -y,z\right)  \Psi_{m}\left(  z\right)
\ dz. \label{green1}%
\end{equation}
Then by (\ref{heis6}) and Fubuni's theorem,%
\begin{align}
&  \sum_{m=0}^{\infty}t_{mn}\left(  \beta\right)  \int_{-\infty}^{\infty
}\left(  -1\right)  ^{m}e^{-imt/2}\ \Psi_{m}\left(  y\right)  \psi_{0}\left(
y\right)  \ dy\label{green2}\\
&  \quad=\sum_{m=0}^{\infty}t_{mn}\left(  \beta\right)  \int_{-\infty}%
^{\infty}\left(  \int_{-\infty}^{\infty}\mathcal{K}_{-t/2}\left(  -y,z\right)
\Psi_{m}\left(  z\right)  \ dz\right)  \psi_{0}\left(  y\right)
\ dy\nonumber\\
&  \quad=\int\int_{-\infty}^{\infty}\mathcal{K}_{-t/2}\left(  -y,z\right)
\left(  \sum_{m=0}^{\infty}t_{mn}\left(  \beta\right)  \Psi_{m}\left(
z\right)  \right)  \psi_{0}\left(  y\right)  \ dydz\nonumber\\
&  \quad=\int\int_{-\infty}^{\infty}\mathcal{K}_{-t/2}\left(  -y,z\right)
\left(  e^{i\beta z}\Psi_{n}\left(  z\right)  \right)  \psi_{0}\left(
y\right)  \ dydz.\nonumber
\end{align}
Now the series (\ref{heis8}) takes the form%
\begin{align}
\psi\left(  x,t\right)   &  =e^{-i\left(  \mu\sin t+\left(  \omega
/2-\mu\right)  t\right)  }\sum_{n=0}^{\infty}\left(  -1\right)  ^{n}%
e^{-it\left(  \omega-1/2\right)  n}\ \Psi_{n}\left(  x\right)  \label{green3}%
\\
&  \quad\times\int\int_{-\infty}^{\infty}\mathcal{K}_{-t/2}\left(
-y,z\right)  \left(  e^{i\beta z}\Psi_{n}\left(  z\right)  \right)  \psi
_{0}\left(  y\right)  \ dydz\nonumber\\
&  =e^{-i\left(  \mu\sin t+\left(  \omega/2-\mu\right)  t\right)  }\int
\int_{-\infty}^{\infty}\mathcal{K}_{-t/2}\left(  -y,z\right)  e^{i\beta
z}\nonumber\\
&  \quad\times\left(  \sum_{n=0}^{\infty}e^{-it\left(  \omega-1/2\right)
n}\ \Psi_{n}\left(  -x\right)  \Psi_{n}\left(  z\right)  \right)  \psi
_{0}\left(  y\right)  \ dydz\nonumber\\
&  =e^{-i\left(  \mu\sin t+\left(  \omega/2-\mu\right)  t\right)  }\nonumber\\
&  \quad\times\int_{-\infty}^{\infty}\left(  \int_{-\infty}^{\infty
}\mathcal{K}_{-t/2}\left(  -y,z\right)  e^{i\beta z}\mathcal{K}_{t\left(
1/2-\omega\right)  }\left(  -x,z\right)  \ dz\right)  \psi_{0}\left(
y\right)  \ dy\nonumber
\end{align}
in view of the generating relations (\ref{four1}) and (\ref{four2}). Thus%
\begin{equation}
\psi\left(  x,t\right)  =e^{-i\left(  \mu\sin t+\left(  \omega/2-\mu\right)
t\right)  }\int_{-\infty}^{\infty}\mathcal{G}_{t}\left(  x,y\right)  \psi
_{0}\left(  y\right)  \ dy, \label{green4}%
\end{equation}
where we define the kernel as%
\begin{equation}
\mathcal{G}_{t}\left(  x,y\right)  :=\int_{-\infty}^{\infty}\mathcal{K}%
_{t\left(  1/2-\omega\right)  }\left(  -x,z\right)  e^{i\beta z}%
\mathcal{K}_{-t/2}\left(  -y,z\right)  \ dz. \label{green5}%
\end{equation}
This can be evaluated with the help of the familiar elementary integrals%
\begin{equation}
\int_{-\infty}^{\infty}e^{-x^{2}}\,dx=\sqrt{\pi},\qquad\int_{-\infty}^{\infty
}e^{i\left(  az^{2}+2bz\right)  }\,dz=\sqrt{\frac{\pi i}{a}}\,e^{-ib^{2}/a}
\label{green6}%
\end{equation}
(see \cite{Bo:Shi}, \cite{ErdInt}, and \cite{Palio:Mead} also). Denoting
$\tau_{1}=t\left(  \omega-1/2\right)  $ and $\tau_{2}=t/2,$ from (\ref{four2})
we get%
\begin{align}
&  \mathcal{K}_{-\tau_{1}}\left(  -x,z\right)  e^{i\beta z}\mathcal{K}%
_{-\tau_{2}}\left(  -y,z\right) \label{green7}\\
&  \quad=\frac{e^{i\left(  \omega t-\pi\right)  /2}}{2\pi\sqrt{\sin\tau
_{1}\sin\tau_{2}}}\ e^{i\left(  x^{2}\cot\tau_{1}+y^{2}\cot\tau_{2}\right)
/2}\ e^{i\left(  \beta+x/\sin\tau_{1}+y/\sin\tau_{2}\right)  z}\ e^{i\left(
\cot\tau_{1}+\cot\tau_{2}\right)  z^{2}/2}\nonumber
\end{align}
and%
\begin{align}
&  \int_{-\infty}^{\infty}\mathcal{K}_{-\tau_{1}}\left(  -x,z\right)
e^{i\beta z}\mathcal{K}_{-\tau_{2}}\left(  -y,z\right)  \ dz\label{green7a}\\
&  \qquad=\frac{e^{i\left(  \omega t-\pi\right)  /2}}{2\pi\sqrt{\sin\tau
_{1}\sin\tau_{2}}}\ e^{i\left(  x^{2}\cot\tau_{1}+y^{2}\cot\tau_{2}\right)
/2}\nonumber\\
&  \qquad\quad\times\int_{-\infty}^{\infty}e^{i\left(  \left(  \beta
+x/\sin\tau_{1}+y/\sin\tau_{2}\right)  z+\left(  \cot\tau_{1}+\cot\tau
_{2}\right)  z^{2}/2\right)  }\ dz.\nonumber
\end{align}
As a result%
\begin{align}
\mathcal{G}_{t}\left(  x,y\right)   &  =\frac{e^{i\left(  \omega
t-\pi/2\right)  /2}}{\sqrt{2\pi\sin\omega t}}\ e^{i\left(  x^{2}\cot\tau
_{1}+y^{2}\cot\tau_{2}\right)  /2}\ \label{green8}\\
&  \quad\times\exp\left(  \frac{\sin\tau_{1}\sin\tau_{2}\left(  \beta
+x/\sin\tau_{1}+y/\sin\tau_{2}\right)  ^{2}}{2i\sin\omega t}\right) \nonumber
\end{align}
with $\tau_{1}=t\left(  \omega-1/2\right)  ,$ $\tau_{2}=t/2$ and $\beta
=\beta\left(  t\right)  =2\sqrt{2\mu}\sin\left(  t/2\right)  .$ Thus the
explicit form of this kernel is given by%
\begin{align}
\mathcal{G}_{t}\left(  x,y\right)   &  =\frac{e^{i\left(  \omega
t-\pi/2\right)  /2}}{\sqrt{2\pi\sin\omega t}}\ \exp\left(  \frac{\left(
x^{2}+y^{2}\right)  \sin\omega t-\left(  x^{2}-y^{2}\right)  \sin\left(
\omega-1\right)  t}{2i\left(  \cos\omega t-\cos\left(  \omega-1\right)
t\right)  }\right) \label{green9}\\
&  \qquad\qquad\qquad\qquad\times\exp\left(  \frac{ik_{t}^{2}\left(
x,y\right)  }{\sin\omega t\left(  \cos\omega t-\cos\left(  \omega-1\right)
t\right)  }\right)  \ ,\nonumber
\end{align}
where%
\begin{align}
k_{t}\left(  x,y\right)   &  =\left(  x+y\right)  \sin\left(  \omega
t/2\right)  \cos\left(  \left(  \omega-1\right)  t/2\right)  -\left(
x-y\right)  \cos\left(  \omega t/2\right)  \sin\left(  \left(  \omega
-1\right)  t/2\right) \label{green10}\\
&  \quad-\sqrt{2\mu}\sin\left(  t/2\right)  \left(  \cos\omega t-\cos\left(
\omega-1\right)  t\right)  .\nonumber
\end{align}
The last expression can be transformed into a somewhat more convenient form%
\begin{align}
&  \mathcal{G}_{t}\left(  x,y\right)  =\mathcal{K}_{\omega t}^{\ast}\left(
x,y\right)  \ \exp\left(  \frac{\sin\left(  \left(  \omega-1/2\right)
t\right)  \ \sin\left(  t/2\right)  \ \beta^{2}}{2i\sin\omega t}\right)
\label{green11}\\
&  \qquad\qquad\quad\times\exp\left(  \frac{\left(  x\sin\left(  t/2\right)
+y\sin\left(  \left(  \omega-1/2\right)  t\right)  \right)  \ \beta}%
{i\sin\omega t}\right) \nonumber
\end{align}
with $\beta=2\sqrt{2\mu}\sin\left(  t/2\right)  $ in terms of the kernel of
the generalized Fourier transform (\ref{four2}). Our formulas (\ref{green4})
and (\ref{green8})--(\ref{green11}) provide an integral form to the solution
of the Cauchy initial value problem (\ref{solu1})--(\ref{solu2}) in terms of a
Green function.\smallskip

By choosing $\psi_{0}\left(  x\right)  =\delta\left(  x-x_{0}\right)  ,$ where
$\delta\left(  x\right)  $ is the Dirac delta function, we formally obtain%
\begin{equation}
\psi\left(  x,t\right)  =G\left(  x,x_{0},t\right)  =e^{-i\left(  \mu\sin
t+\left(  \omega/2-\mu\right)  t\right)  }\mathcal{G}_{t}\left(
x,x_{0}\right)  , \label{green12}%
\end{equation}
which is the fundamental solution to the time-dependent Schr\"{o}dinger
equation (\ref{solu1}). One can show that%
\begin{equation}
\lim_{t\rightarrow0^{+}}\psi\left(  x,t\right)  =\psi_{0}\left(  x\right)
\label{green13}%
\end{equation}
by methods of \cite{As:Ra:Su}, \cite{Ra:Su} and \cite{Wi}. The details are
left to the reader.\smallskip

The time evolution operator for the time-dependent Schr\"{o}dinger equation
(\ref{int6}) can formally be written as%
\begin{equation}
U\left(  t,t_{0}\right)  =\text{T}\left(  \exp\left(  -\frac{i}{\hslash}%
\int_{t_{0}}^{t}H\left(  t^{\prime}\right)  \ dt^{\prime}\right)  \right)  ,
\label{green14}%
\end{equation}
where T is the time ordering operator which orders operators with larger times
to the left \cite{Bo:Shi}, \cite{Flu}. Namely, this unitary operator takes a
state at time $t_{0}$ to a state at time $t,$ so that%
\begin{equation}
\psi\left(  x,t\right)  =U\left(  t,t_{0}\right)  \psi\left(  x,t_{0}\right)
\label{green15}%
\end{equation}
and%
\begin{equation}
U\left(  t,t_{0}\right)  =U\left(  t,t^{\prime}\right)  U\left(  t^{\prime
},t_{0}\right)  , \label{green14a}%
\end{equation}%
\begin{equation}
U^{-1}\left(  t,t_{0}\right)  =U^{\dagger}\left(  t,t_{0}\right)  =U\left(
t_{0},t\right)  . \label{green15a}%
\end{equation}
We have constructed this time evolution operator explicitly as the following
integral operator%
\begin{equation}
U\left(  t,t_{0}\right)  \psi\left(  x,t_{0}\right)  =e^{-i\left(  \mu
\sin\left(  t-t_{0}\right)  +\left(  \omega/2-\mu\right)  \left(
t-t_{0}\right)  \right)  }\int_{-\infty}^{\infty}\mathcal{G}_{t-t_{0}}\left(
x,y\right)  \psi\left(  y,t_{0}\right)  \ dy \label{green16}%
\end{equation}
with the kernel given by (\ref{green8})--(\ref{green11}), for the particular
form of the time-dependent Hamiltonian in (\ref{harm1})--(\ref{harm3}) and
(\ref{separ6}). The Green function (propagator) for the general forced
harmonic oscillator is constructed in section~10 ; see equations
(\ref{fho3})--(\ref{fho8}).

\section{The Cauchy Problem for the Simple Harmonic Oscillator}

In an important special case $\mu=0,$ the initial value problem%
\begin{equation}
i\frac{\partial\psi}{\partial t}=\frac{\omega}{2}\left(  -\frac{\partial
^{2}\psi}{\partial x^{2}}+x^{2}\psi\right)  ,\qquad\left.  \psi\left(
x,t\right)  \right\vert _{t=0}=\psi_{0}\left(  x\right)  \quad\left(
-\infty<x<\infty\right)  \label{cpharm1}%
\end{equation}
has the following explicit solution%
\begin{align}
\psi\left(  x,t\right)   &  =\sum_{n=0}^{\infty}e^{-i\omega\left(
n+1/2\right)  t}\ \Psi_{n}\left(  x\right)  \int_{-\infty}^{\infty}\ \Psi
_{n}\left(  y\right)  \psi_{0}\left(  y\right)  \ dy\label{cpharm2}\\
&  =\frac{1}{\sqrt{2\pi i\sin\left(  \omega t\right)  }}\
%TCIMACRO{\dint _{-\infty}^{\infty}}%
%BeginExpansion
{\displaystyle\int_{-\infty}^{\infty}}
%EndExpansion
\exp\left(  i\dfrac{\left(  x^{2}+y^{2}\right)  \cos\left(  \omega t\right)
-2xy}{2\sin\left(  \omega t\right)  }\right)  \ \psi_{0}\left(  y\right)
\ dy.\nonumber
\end{align}
The last relation is valid when $0<t<\pi/\omega.$ Analytic continuation in a
larger domain is discussed in \cite{Maslov:Fedoriuk} and \cite{Thomber:Taylor}%
.\smallskip\ 

Equation (\ref{cpharm2}) gives the time evolution operator (\ref{green14}) for
the simple harmonic oscillator in terms of the generalized Fourier transform.
This result and its extension to a general forced harmonic oscillator without
the velocity-dependent term in the Hamiltonian are well-known (see
\cite{Beauregard}, \cite{Feynman}, \cite{Fey:Hib}, \cite{Gottf:T-MY},
\cite{Holstein}, \cite{Maslov:Fedoriuk}, \cite{Merz}, \cite{Thomber:Taylor}
and references therein; further generalizations are given in sections~10--12;
more special cases will be discussed in section~13).

\section{Three Parameter Generalization of the Fourier Transform}

The properties of the time evolution operator in (\ref{green15}%
)--(\ref{green16}) suggest the following extension of the classical Fourier
integral%
\begin{equation}
f\left(  x\right)  =\int_{-\infty}^{\infty}\mathcal{L}_{t}\left(  x,y\right)
g\left(  y\right)  \ dy, \label{genfour1}%
\end{equation}
where the kernel given by%
\begin{align}
\mathcal{L}_{t}\left(  x,y\right)   &  =\mathcal{K}_{\omega t}\left(
x,y\right)  \ \exp\left(  i\frac{\sin\left(  \left(  \omega-1/2\right)
t\right)  \ \sin^{3}\left(  t/2\right)  }{2\sin\omega t}\ \varepsilon
^{2}\right) \label{genfour2}\\
&  \times\exp\left(  i\frac{\left(  x\sin\left(  t/2\right)  +y\sin\left(
\left(  \omega-1/2\right)  t\right)  \right)  \sin\left(  t/2\right)  }%
{\sin\omega t}\ \varepsilon\right) \nonumber
\end{align}
depends on the three free parameters $t,$ $\omega$ and $\varepsilon.$ If
$\varepsilon=0$ and $\omega t=\tau$ we arrive at the kernel of the generalized
Fourier transform (\ref{four2}). The formal inversion formula is given by%
\begin{equation}
g\left(  y\right)  =\int_{-\infty}^{\infty}\mathcal{L}_{t}^{\ast}\left(
x,y\right)  f\left(  x\right)  \ dx. \label{genfour3}%
\end{equation}
The details are left to the reader. Note that, in terms of a distribution,%
\begin{align}
&  \int_{-\infty}^{\infty}\mathcal{L}_{t}^{\ast}\left(  x,y\right)
\mathcal{L}_{t}\left(  x,z\right)  \ dx\label{genfour4}\\
&  \qquad=e^{i\cot\left(  \omega t\right)  \left(  y^{2}-z^{2}\right)
/2}\ e^{i\varepsilon\left(  z-y\right)  \sin\left(  \left(  \omega-1/2\right)
t\right)  \sin\left(  t/2\right)  /\sin\omega t}\qquad\quad\nonumber\\
&  \quad\qquad\times\frac{1}{2\pi\sin\omega t}\int_{-\infty}^{\infty
}e^{ix\left(  z-y\right)  /\sin\omega t}\ dx=\delta\left(  y-z\right)
,\nonumber
\end{align}
which gives the corresponding orthogonality property of the $\mathcal{L}%
$-kernel. These results admit further generalizations with the help of the
time evolution operators found in sections~10 and 12.

\section{The General Forced Harmonic Oscillator}

Our solution of the initial value problem (\ref{solu1})--(\ref{solu2})
obtained in the previous sections admits a generalization. The Cauchy problem
for the general forced harmonic oscillator%
\begin{equation}
i\frac{\partial\psi}{\partial t}=\frac{\omega}{2}\left(  -\frac{\partial
^{2}\psi}{\partial x^{2}}+x^{2}\psi\right)  -f\left(  t\right)  x\ \psi
+ig\left(  t\right)  \ \frac{\partial\psi}{\partial x}, \label{fho1}%
\end{equation}
where $f\left(  t\right)  $ and $g\left(  t\right)  $ are two arbitrary real
valued functions of time only (such that the integrals in (\ref{fho6}%
)--(\ref{fho8}) below converge and $a\left(  0\right)  =0$), with the initial
data%
\begin{equation}
\left.  \psi\left(  x,t\right)  \right\vert _{t=0}=\psi_{0}\left(  x\right)
\qquad\left(  -\infty<x<\infty\right)  \label{fho2}%
\end{equation}
has the following explicit solution%
\begin{equation}
\psi\left(  x,t\right)  =\int_{-\infty}^{\infty}G\left(  x,y,t\right)
\psi_{0}\left(  y\right)  \ dy. \label{fho3}%
\end{equation}
Here the Green function (or Feynman's propagator \cite{Feynman},
\cite{Fey:Hib}, \cite{Merz}) is given by%
\begin{equation}
G\left(  x,y,t\right)  =G_{0}\left(  x,y,t\right)  \ e^{i\left(  a\left(
t\right)  x+b\left(  t\right)  y+c\left(  t\right)  \right)  } \label{fho4}%
\end{equation}
with%
\begin{equation}
G_{0}\left(  x,y,t\right)  =\frac{1}{\sqrt{2\pi i\sin\omega t}}\ \exp\left(
i\dfrac{\left(  x^{2}+y^{2}\right)  \cos\omega t-2xy}{2\sin\omega t}\right)
\label{fho5}%
\end{equation}
and%
\begin{align}
a\left(  t\right)   &  =\frac{1}{\sin\omega t}\int_{0}^{t}\left(  f\left(
s\right)  \sin\omega s+g\left(  s\right)  \cos\omega s\right)
\ ds,\label{fho6}\\
b\left(  t\right)   &  =\int_{0}^{t}\frac{\omega a\left(  s\right)  -g\left(
s\right)  }{\sin\omega s}\ ds,\label{fho7}\\
c\left(  t\right)   &  =\int_{0}^{t}\left(  g\left(  s\right)  a\left(
s\right)  -\frac{\omega}{2}a^{2}\left(  s\right)  \right)  \ ds \label{fho8}%
\end{align}
provided $a\left(  0\right)  =b\left(  0\right)  =c\left(  0\right)  =0.$ The
case $g\left(  t\right)  \equiv0$ is discussed in \cite{Feynman},
\cite{Fey:Hib} and \cite{Merz} but the answers for $b\left(  t\right)  $ and
$c\left(  t\right)  $ are given in different forms; we shall elaborate on this
later.\smallskip

Indeed, the previously found solution (\ref{green11})--(\ref{green12}) in the
special case of the forced oscillator (\ref{solu1}) suggests to look for a
general Green function in the form (\ref{fho4}), namely,%
\begin{equation}
\psi=u\ e^{iS}, \label{fho4a}%
\end{equation}
where $u=G_{0}\left(  x,y,t\right)  $ is the fundamental solution of the
Sch\"{o}dinger equation for the simple harmonic oscillator (\ref{cpharm1}) and
$S=a\left(  t\right)  x+b\left(  t\right)  y+c\left(  t\right)  .$ Its
substitution into (\ref{fho1}) gives%
\begin{equation}
\left(  \frac{da}{dt}x+\frac{db}{dt}y+\frac{dc}{dt}\right)  u=\left(
ag+xf-\frac{\omega}{2}a^{2}\right)  u+i\left(  a\omega-g\right)
\frac{\partial u}{\partial x}, \label{fho4b}%
\end{equation}
where by (\ref{fho5})%
\begin{equation}
\frac{\partial u}{\partial x}=i\frac{x\cos\omega t-y}{\sin\omega t}\ u.
\label{fho5a}%
\end{equation}
As a result%
\begin{equation}
\frac{da}{dt}x+\frac{db}{dt}y+\frac{dc}{dt}=ag+xf-\frac{\omega}{2}%
a^{2}-\left(  a\omega-g\right)  \frac{x\cos\omega t-y}{\sin\omega t},
\label{fho4ab}%
\end{equation}
and equating the coefficients of $x,$ $y$ and $1,$ we obtain the following
system of ordinary differential equations%
\begin{align}
\frac{d}{dt}\left(  \sin\omega t\ a\left(  t\right)  \right)   &  =f\left(
t\right)  \sin\omega t+g\left(  t\right)  \cos\omega t,\label{fho9}\\
\frac{d}{dt}b\left(  t\right)   &  =\frac{\omega a\left(  t\right)  -g\left(
t\right)  }{\sin\omega t},\label{fho10}\\
\frac{d}{dt}c\left(  t\right)   &  =g\left(  t\right)  a\left(  t\right)
-\frac{\omega}{2}a^{2}\left(  t\right)  , \label{fho11}%
\end{align}
whose solutions are (\ref{fho6})--(\ref{fho8}), respectively, if the integrals
converge. This method is equivalent to solving of the quantum mechanical
Hamilton--Jacobi equation for the general forced harmonic oscillator
\cite{Merz}.\smallskip

Equation (\ref{fho7}) can be rewritten as%
\[
b\left(  t\right)  =-\int_{0}^{t}\left(  \sin\omega s\ a\left(  s\right)
\right)  \ d\cot\omega s-\int_{0}^{t}\frac{g\left(  s\right)  }{\sin\omega
s}\ ds
\]
and integrating by parts%
\begin{equation}
b\left(  t\right)  =-\cos\omega t\ a\left(  t\right)  +\int_{0}^{t}\left(
f\left(  s\right)  \cos\omega s-g\left(  s\right)  \sin\omega s\right)  \ ds
\label{fho7a}%
\end{equation}
by (\ref{fho9}). With the help of (\ref{fho6}) and the addition formulas for
trigonometric functions we finally arrive at%
\begin{equation}
b\left(  t\right)  =-\frac{1}{\sin\omega t}\int_{0}^{t}\left(  f\left(
s\right)  \sin\omega\left(  s-t\right)  +g\left(  s\right)  \cos\omega\left(
s-t\right)  \right)  \ ds, \label{fho7b}%
\end{equation}
which is equivalent to the form obtain in \cite{Feynman}, \cite{Fey:Hib} and
\cite{Merz} when $g\left(  t\right)  \equiv0.$\smallskip\ 

In a similar fashion,%
\[
c\left(  t\right)  =\int_{0}^{t}g\left(  s\right)  a\left(  s\right)
\ ds+\frac{1}{2}\int_{0}^{t}\left(  \sin\omega s\ a\left(  s\right)  \right)
^{2}\ d\cot\omega s
\]
and as a result%
\begin{equation}
c\left(  t\right)  =\frac{1}{2}\sin\omega t\cos\omega t\ a^{2}\left(
t\right)  +\int_{0}^{t}\sin\omega s\ a\left(  s\right)  \left(  -f\left(
s\right)  \cos\omega s+g\left(  s\right)  \sin\omega s\right)  \ ds.
\label{fho8a}%
\end{equation}
This can be transformed into the form given in \cite{Feynman} and
\cite{Fey:Hib} when $g\left(  t\right)  \equiv0.$ The details are left to the
reader.\smallskip\ 

Evaluation of elementary integrals results in (\ref{green11}) again in the
special case (\ref{solu1}). The simple case $f\left(  t\right)  =2\cos\omega
t$ and $g\left(  t\right)  \equiv0$ gives%
\begin{equation}
a\left(  t\right)  =\frac{\sin\omega t}{\omega},\quad\quad b\left(  t\right)
=t,\quad\quad c\left(  t\right)  =\frac{1}{8\omega^{2}}\sin2\omega t-\frac
{1}{4\omega}t. \label{fho12}%
\end{equation}
The corresponding propagator in (\ref{fho4}) does satisfy the Schr\"{o}dinger
equation (\ref{fho1}), which can be verified by a direct differentiation with
the help of a computer algebra system. The details are left to the reader. A
case of the forced modified oscillator is discussed in \cite{Me:Co:Su}.

\section{Eigenfunction Expansion for the General Forced Harmonic Oscillator}

Separation of the $x$ and $y$ variables in Feynman's propagator (\ref{fho4}%
)--(\ref{fho8}) with the help of the Mehler generating function (\ref{four1})
written as%
\begin{equation}
G_{0}\left(  x,y,t\right)  =\sum_{k=0}^{\infty}e^{-i\omega\left(
k+1/2\right)  t}\ \Psi_{k}\left(  x\right)  \Psi_{k}\left(  y\right)
\label{effho1}%
\end{equation}
gives%
\begin{align}
G\left(  x,y,t\right)   &  =G_{0}\left(  x,y,t\right)  \ e^{i\left(
ax+by+c\right)  }\label{effho2}\\
&  =e^{i\left(  c-\omega t/2\right)  }\sum_{k=0}^{\infty}e^{-i\omega
kt}\ \left(  e^{iax}\Psi_{k}\left(  x\right)  \right)  \left(  e^{iby}\Psi
_{k}\left(  y\right)  \right) \nonumber\\
&  =e^{i\left(  c-\omega t/2\right)  }\sum_{k=0}^{\infty}e^{-i\omega
kt}\ \left(  \sum_{n=0}^{\infty}t_{nk}\left(  a\right)  \ \Psi_{n}\left(
x\right)  \right)  \left(  \sum_{m=0}^{\infty}t_{mk}\left(  b\right)
\ \Psi_{m}\left(  y\right)  \right) \nonumber\\
&  =e^{i\left(  c-\omega t/2\right)  }\sum_{n=0}^{\infty}\sum_{m=0}^{\infty
}\Psi_{n}\left(  x\right)  \Psi_{m}\left(  y\right)  \left(  \sum
_{k=0}^{\infty}e^{-i\omega kt}\ t_{nk}\left(  a\right)  t_{mk}\left(
b\right)  \right) \nonumber
\end{align}
by (\ref{heis6}). The last series can be summed by using the addition formula
(\ref{heis6a}) in the form%
\begin{align}
\sum_{k=0}^{\infty}e^{-i\omega kt}\ t_{nk}\left(  a\right)  t_{mk}\left(
b\right)   &  =\frac{i^{m+n}}{\sqrt{2^{m+n}m!n!}}\ e^{i\left(  ab\sin\omega
t\right)  /2}\ e^{-\chi^{2}/4}\label{effho3}\\
&  \quad\times\left(  a+bz\right)  ^{n}\left(  b+az\right)  ^{m}\ c_{m}%
^{\chi^{2}/2}\left(  n\right)  ,\nonumber
\end{align}
with $z=e^{-i\omega t}$ and $\chi^{2}=a^{2}+b^{2}+2ab\cos\omega t.$ As a
result we arrive at the following eigenfunction expansion of the forced
harmonic oscillator propagator%
\begin{align}
&  G\left(  x,y,t\right)  =e^{i\left(  c-\left(  \omega t-ab\sin\omega
t\right)  /2\right)  }\ e^{-\chi^{2}/4}\label{effho4}\\
&  \quad\times\sum_{n=0}^{\infty}\sum_{m=0}^{\infty}\Psi_{n}\left(  x\right)
\Psi_{m}\left(  y\right)  \ \frac{i^{n+m}}{\sqrt{2^{n+m}n!m!}}\ \left(
a+bz\right)  ^{n}\left(  b+az\right)  ^{m}c_{m}^{\chi^{2}/2}\left(  n\right)
\nonumber
\end{align}
in terms of the Charlier polynomials. The special case $g\left(  t\right)
\equiv0$ is discussed in \cite{Fey:Hib} but the connection with the Charlier
polynomials is not emphasized.\smallskip

The solution (\ref{fho3}) takes the form%
\begin{equation}
\psi\left(  x,t\right)  =\sum_{n=0}^{\infty}\Psi_{n}\left(  x\right)
\sum_{m=0}^{\infty}c_{nm}\left(  t\right)  \ \int_{-\infty}^{\infty}\Psi
_{m}\left(  y\right)  \psi_{0}\left(  y\right)  \ dy, \label{effho5}%
\end{equation}
where%
\begin{equation}
c_{nm}\left(  t\right)  =e^{i\left(  c-\left(  \omega t-ab\sin\omega t\right)
/2\right)  }\ e^{-\chi^{2}/4}\frac{i^{n+m}}{\sqrt{2^{n+m}n!m!}}\ \left(
a+bz\right)  ^{n}\left(  b+az\right)  ^{m}c_{m}^{\chi^{2}/2}\left(  n\right)
\label{effho6}%
\end{equation}
with $z=e^{-i\omega t}$ and $\chi^{2}=a^{2}+b^{2}+2ab\cos\omega t.$ Functions
$a=a\left(  t\right)  ,$ $b=b\left(  t\right)  $ and $c=c\left(  t\right)  $
here are given by the integrals (\ref{fho6})--(\ref{fho8}), respectively, and
$\lim_{t\rightarrow0^{+}}c_{nm}\left(  t\right)  =\delta_{nm}.$\smallskip

If $\psi_{0}\left(  x\right)  =\left.  \psi\left(  x,t\right)  \right\vert
_{t=0}=\Psi_{m}\left(  x\right)  ,$ equation (\ref{effho5}) becomes%
\begin{equation}
\psi\left(  x,t\right)  =\sum_{n=0}^{\infty}c_{nm}\left(  t\right)  \ \Psi
_{n}\left(  x\right)  . \label{effho7}%
\end{equation}
Thus function $c_{nm}\left(  t\right)  $ gives explicitly the quantum
mechanical amplitude that the oscillator initially in state $m$ is found at
time $t$ in state $n$ \cite{Fey:Hib}. An application to the motion of a
charged particle with a spin in uniform perpendicular magnetic and electric
fields is considered in section~14. \smallskip

As a by-product we found the fundamental solution $c_{nm}\left(  t\right)  $
of the system (\ref{separ2}) in terms of the Charlier polynomials for an
arbitrary complex valued function $\delta\left(  t\right)  =\left(  -f\left(
t\right)  +ig\left(  t\right)  \right)  /\sqrt{2}.$ The explicit solution of
the corresponding initial value problem in given by (\ref{separ7}).

\section{Time-Dependent Frequency}

An extension of the Sch\"{o}dinger equation to the case of the forced harmonic
oscillator with the time-dependent frequency is as follows%
\begin{equation}
i\frac{\partial\psi}{\partial t}=\frac{\omega\left(  t\right)  }{2}\left(
-\frac{\partial^{2}\psi}{\partial x^{2}}+x^{2}\psi\right)  -f\left(  t\right)
x\ \psi+ig\left(  t\right)  \ \frac{\partial\psi}{\partial x}, \label{omt1}%
\end{equation}
where $\omega\left(  t\right)  >0,$ $f\left(  t\right)  $ and $g\left(
t\right)  $ are arbitrary real valued functions of time only. It can be easily
solved by the substitution%
\begin{equation}
\tau=\tau\left(  t\right)  =\int_{0}^{t}\omega\left(  s\right)  \ ds,\qquad
\frac{d\tau}{dt}=\omega\left(  t\right)  , \label{omt2}%
\end{equation}
which transforms this equation into a familiar form%
\begin{equation}
i\frac{\partial\psi}{\partial\tau}=\frac{1}{2}\left(  -\frac{\partial^{2}\psi
}{\partial x^{2}}+x^{2}\psi\right)  -f_{1}\left(  \tau\right)  x\ \psi
+ig_{1}\left(  \tau\right)  \ \frac{\partial\psi}{\partial x}, \label{omt3}%
\end{equation}
see the original equation (\ref{fho1}) with respect to the new time variable
$\tau,$ where $\omega=1$ and
\begin{equation}
f_{1}\left(  \tau\right)  =\frac{f\left(  t\right)  }{\omega\left(  t\right)
},\qquad g_{1}\left(  \tau\right)  =\frac{f\left(  t\right)  }{\omega\left(
t\right)  }. \label{omt4}%
\end{equation}
Therefore by (\ref{fho4})--(\ref{fho5}) the propagator has the form%
\begin{equation}
G\left(  x,y,\tau\right)  =G_{0}\left(  x,y,\tau\right)  \ e^{i\left(
a\left(  \tau\right)  x+b\left(  \tau\right)  y+c\left(  \tau\right)  \right)
} \label{omt5}%
\end{equation}
with%
\begin{equation}
G_{0}\left(  x,y,\tau\right)  =\frac{1}{\sqrt{2\pi i\sin\tau}}\ \exp\left(
i\dfrac{\left(  x^{2}+y^{2}\right)  \cos\tau-2xy}{2\sin\tau}\right)
\label{omt6}%
\end{equation}
and the system (\ref{fho9})--(\ref{fho11}) becomes%
\begin{align}
\frac{d}{d\tau}\left(  \sin\tau\ a\left(  \tau\right)  \right)   &
=f_{1}\left(  \tau\right)  \sin\tau+g_{1}\left(  \tau\right)  \cos
\tau,\label{omt7}\\
\frac{d}{d\tau}b\left(  \tau\right)   &  =\frac{a\left(  \tau\right)
-g_{1}\left(  \tau\right)  }{\sin\tau},\label{omt8}\\
\frac{d}{d\tau}c\left(  \tau\right)   &  =g_{1}\left(  \tau\right)  a\left(
\tau\right)  -\frac{1}{2}a^{2}\left(  \tau\right)  . \label{omt9}%
\end{align}
Thus%
\begin{align}
a\left(  \tau\right)   &  =\frac{1}{\sin\tau}\ \int_{0}^{t}\left(  f\left(
s\right)  \sin\tau\left(  s\right)  +g\left(  s\right)  \cos\tau\left(
s\right)  \right)  \ ds,\label{omt10}\\
b\left(  \tau\right)   &  =\int_{0}^{t}\frac{\omega\left(  s\right)  a\left(
\tau\left(  s\right)  \right)  -g\left(  s\right)  }{\sin\tau\left(  s\right)
}\ ds,\label{omt11}\\
c\left(  \tau\right)   &  =\int_{0}^{t}\left(  g\left(  s\right)  a\left(
\tau\left(  s\right)  \right)  -\frac{1}{2}\omega\left(  s\right)
a^{2}\left(  \tau\left(  s\right)  \right)  \right)  \ ds, \label{omt12}%
\end{align}
which is an extension of equations (\ref{fho6})--(\ref{fho7}) to the case of
the forced harmonic oscillator with the time-dependent frequency. The solution
of the Cauchy initial value problem is given by%
\begin{equation}
\psi\left(  x,t\right)  =\int_{-\infty}^{\infty}G\left(  x,y,\tau\right)
\psi_{0}\left(  y\right)  \ dy \label{omt13}%
\end{equation}
with $\tau=%
%TCIMACRO{\dint _{0}^{t}}%
%BeginExpansion
{\displaystyle\int_{0}^{t}}
%EndExpansion
\omega\left(  s\right)  \ ds.$ The details are left to the reader.

\section{Some Special Cases}

The time-dependent Schr\"{o}dinger equation for the forced harmonic oscillator
is usually written in the form%
\begin{equation}
i\hslash\frac{\partial\Psi}{\partial t}=H\Psi\label{spc1}%
\end{equation}
with the following Hamiltonian%
\begin{equation}
H=\frac{p^{2}}{2m}+\frac{m\omega^{2}}{2}x^{2}-F\left(  t\right)  x-G\left(
t\right)  p,\qquad p=\frac{\hslash}{i}\frac{\partial}{\partial x},
\label{spc2}%
\end{equation}
where $\hslash$ is the Planck constant, $m$ is the mass of the particle,
$\omega$ is the classical oscillation frequency, $F\left(  t\right)  $ is a
uniform in space external force depending on time, function $G\left(
t\right)  $ represents a similar velocity-dependent term and $p$ is the linear
momentum operator. The initial value problem is%
\begin{equation}
i\hslash\frac{\partial\Psi}{\partial t}=-\frac{\hslash^{2}}{2m}\frac
{\partial^{2}\Psi}{\partial x^{2}}+\frac{m\omega^{2}}{2}x^{2}\ \Psi-F\left(
t\right)  x\ \Psi+i\hslash G\left(  t\right)  \ \frac{\partial\Psi}{\partial
x} \label{spc3}%
\end{equation}
with%
\begin{equation}
\left.  \Psi\left(  x,t\right)  \right\vert _{t=t_{0}}=\Psi\left(
x,t_{0}\right)  . \label{spc4}%
\end{equation}
Among important special cases are: the free particle, when $\omega=F=G=0;$ a
particle in a constant external field, where $\omega=G=0$ and $F=\ $constant;
the simple harmonic oscillator with $F=G=0.$ In this section for the benefits
of the reader we provide explicit forms for the corresponding propagators by
taking certain limits in the general solution.\smallskip

The usual change of the space variable%
\begin{equation}
\Psi\left(  x,t\right)  =\psi\left(  \xi,t\right)  ,\qquad\xi=\sqrt
{\frac{m\omega}{\hslash}}\ x \label{spc5}%
\end{equation}
reduces equation (\ref{spc3}) to the form (\ref{fho1}) with respect to $\xi$
with%
\begin{equation}
f\left(  t\right)  =\frac{F\left(  t\right)  }{\sqrt{\hslash\omega m}},\qquad
g\left(  t\right)  =\sqrt{\frac{m\omega}{\hslash}}\ G\left(  t\right)  .
\label{spc6}%
\end{equation}
The time evolution operator is%
\begin{equation}
\Psi\left(  x,t\right)  =\int_{-\infty}^{\infty}G\left(  x,y,t,t_{0}\right)
\ \Psi\left(  y,t_{0}\right)  \ dy \label{spc7}%
\end{equation}
with the propagator of the form%
\begin{equation}
G\left(  x,y,t,t_{0}\right)  =G_{0}\left(  x,y,t,t_{0}\right)  \ e^{i\left(
a\left(  t,t_{0}\right)  x+b\left(  t,t_{0}\right)  y+c\left(  t,t_{0}\right)
\right)  }. \label{spc8}%
\end{equation}
Here%
\begin{align}
G_{0}\left(  x,y,t,t_{0}\right)   &  =\sqrt{\frac{m\omega}{2\pi i\hslash
\sin\omega\left(  t-t_{0}\right)  }}\ \label{spc9}\\
&  \quad\times\exp\left(  \dfrac{im\omega}{2\hslash\sin\omega\left(
t-t_{0}\right)  }\left(  \left(  x^{2}+y^{2}\right)  \cos\omega\left(
t-t_{0}\right)  -2xy\right)  \right)  ,\nonumber
\end{align}%
\begin{align}
a\left(  t,t_{0}\right)   &  =\frac{m\omega}{\hslash\sin\omega\left(
t-t_{0}\right)  }\ \int_{t_{0}}^{t}\left(  F\left(  s\right)  \frac{\sin
\omega\left(  s-t_{0}\right)  }{m\omega}+G\left(  s\right)  \cos\omega\left(
s-t_{0}\right)  \right)  \ ds,\label{spc10}\\
b\left(  t,t_{0}\right)   &  =-a\left(  t_{0},t\right)  \label{spc11}%
\end{align}
and%
\begin{equation}
c\left(  t,t_{0}\right)  =\int_{t_{0}}^{t}\left(  G\left(  s\right)
\ a\left(  s,t_{0}\right)  -\frac{\hslash}{2m}\ a^{2}\left(  s,t_{0}\right)
\right)  \ ds. \label{spc12}%
\end{equation}

The simple harmonic oscillator propagator, when $F=G=0,$ is given by equation
(\ref{spc9}); see \cite{Beauregard}, \cite{Feynman}, \cite{Fey:Hib},
\cite{Gottf:T-MY}, \cite{Holstein}, \cite{Maslov:Fedoriuk}, \cite{Merz},
\cite{Thomber:Taylor} and references therein for more details. In the limit
$\omega\rightarrow0$ we obtain%
\begin{equation}
G_{0}\left(  x,y,t,t_{0}\right)  =\sqrt{\frac{m}{2\pi i\hslash\left(
t-t_{0}\right)  }}\ \exp\left(  \frac{im\left(  x-y\right)  ^{2}}%
{2\hslash\left(  t-t_{0}\right)  }\right)  \label{spc13}%
\end{equation}
as the free particle propagator \cite{Fey:Hib}.\smallskip

For a particle in a constant external field $\omega=G=0$ and $F=\ $constant.
The corresponding propagator is given by%
\begin{align}
G\left(  x,y,t,t_{0}\right)   &  =\sqrt{\frac{m}{2\pi i\hslash\left(
t-t_{0}\right)  }}\ \exp\left(  \frac{im\left(  x-y\right)  ^{2}}%
{2\hslash\left(  t-t_{0}\right)  }\right) \label{spc14}\\
&  \quad\times\exp\left(  \frac{iF\left(  x+y\right)  }{2\hslash}\left(
t-t_{0}\right)  -\frac{iF^{2}}{24\hslash m}\left(  t-t_{0}\right)
^{3}\right)  .\nonumber
\end{align}
This case was studied in detail in \cite{Arrighini:Durante},
\cite{Brown:Zhang}, \cite{Fey:Hib}, \cite{Holstein97}, \cite{Nardone} and
\cite{Robinett}. We have corrected a typo in \cite{Fey:Hib}.

\section{Motion in Uniform Perpendicular Magnetic and Electric Fields}

%%%%%%%%%%%%%%%%%%%%%%%%%%%%%%%%%%%%%%%%%%%
%%%This WinEdt version of the figure 1%%%
\begin{figure}[ptbh]
\centering\scalebox{.65}{\includegraphics{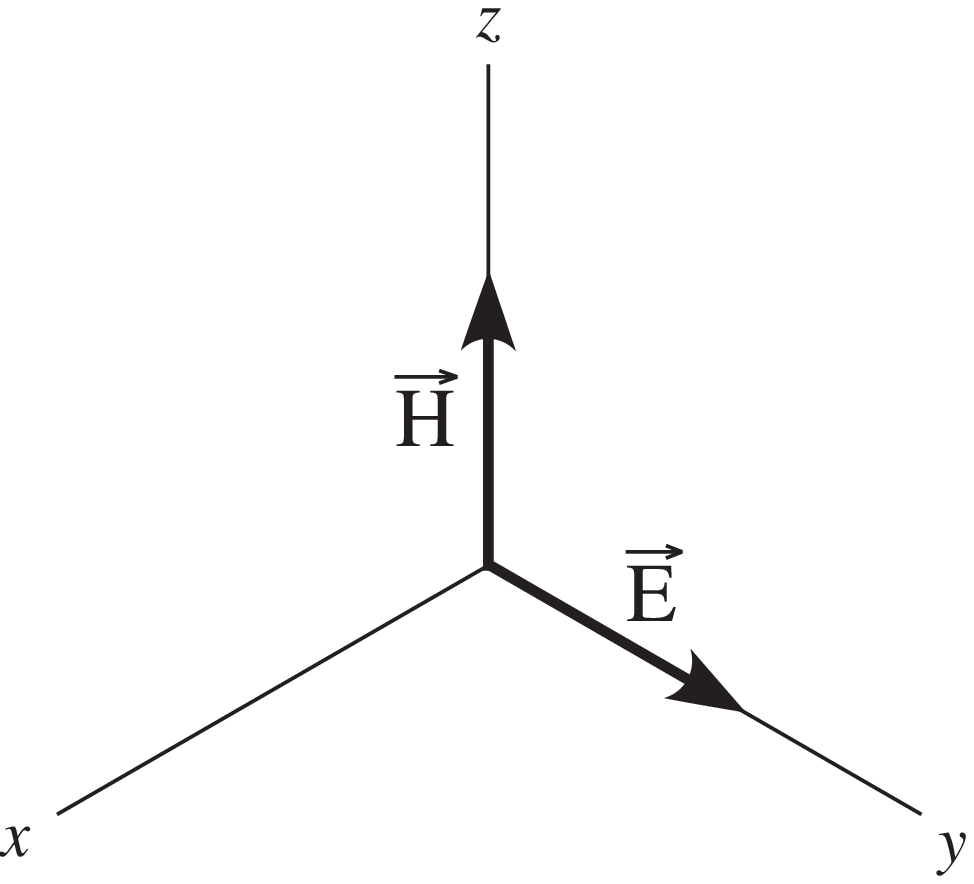}}\caption{Magnetic and
electric fields in $\boldsymbol{R}^{3}$.}%
\end{figure}
%%%%%%%%%%%%%%%%%%%%%%%%%%%%%%%%%%%%%%%%%%%

\subsection{Solution of a Particular Initial Value Problem}

A particle with a spin $s$ has also an intrinsic magnetic momentum
$\boldsymbol{\mu}$ with the operator%
\begin{equation}
\widehat{\boldsymbol{\mu}}=\mu\widehat{\boldsymbol{s}}/s,\label{ll1}%
\end{equation}
where $\widehat{\boldsymbol{s}}$ is the spin operator and $\mu$ is a constant
characterizing the particle, which is usually called the magnitude of the
magnetic momentum. For the motion of a charged particle in uniform magnetic
$\boldsymbol{H}$ and electric $\boldsymbol{E}$ fields, which are perpendicular
to each other (Figure~1), the corresponding three-dimensional time-dependent
Schr\"{o}dinger equation%
\begin{equation}
i\hslash\frac{\partial\Psi}{\partial t}=\widehat{H}\Psi\label{ll2}%
\end{equation}
has the Hamiltonian of the form \cite{La:Lif}%
\begin{equation}
\widehat{H}=\frac{1}{2m}\left(  \widehat{p}_{x}+\frac{eH}{c}y\right)
^{2}+\frac{1}{2m}\widehat{p}_{y}^{2}+\frac{1}{2m}\widehat{p}_{z}^{2}-\frac
{\mu}{s}\widehat{s}_{z}H-yF,\label{ll3}%
\end{equation}
where $\widehat{\boldsymbol{p}}=-i\hslash\nabla$ is the linear momentum
operator, functions $H$ and $F/e$ are the magnitudes of the uniform magnetic
and electric fields in $z$ and $y$ directions, respectively. The corresponding
vector potential $\boldsymbol{A}=-yH\ \boldsymbol{e}_{x}$ is defined up to a
gauge transformation. Here we follow the original choice of \cite{La:Lif} (see
a remark at the end of this section).\smallskip

Since (\ref{ll3}) does not contain the other components of the spin, the
operator $\widehat{s}_{z}$ commutes with the Hamiltonian $\widehat{H}$ and the
$z$-component of the spin is conserved. Thus the operator $\widehat{s}_{z}$
can be replaced by its eigenvalue $s_{z}=\sigma$ in the Hamiltonian%
\begin{equation}
\widehat{H}=\frac{1}{2m}\left(  \widehat{p}_{x}+\frac{eH}{c}y\right)
^{2}+\frac{1}{2m}\widehat{p}_{y}^{2}+\frac{1}{2m}\widehat{p}_{z}^{2}-\frac
{\mu\sigma}{s}H-yF \label{ll4}%
\end{equation}
with $\sigma=-s,-s+1,...\ ,s-1,s.$ Then the spin dependence of the wave
function becomes unimportant and the wave function in the Schr\"{o}dinger
equation (\ref{ll2}) can be taken as an ordinary coordinate function
$\Psi=\Psi\left(  \boldsymbol{r},t,\sigma\right)  .$\smallskip

The Hamiltonian (\ref{ll4}) does not contain the coordinates $x$ and $z$
explicitly. Therefore the operators $\widehat{p}_{x}$ and $\widehat{p}_{z}$
also commute with the Hamiltonian and the $x$ and $z$ components of the linear
momentum are conserved. The corresponding eigenvalues $p_{x}$ and $p_{z}$ take
all values from $-\infty$ to $\infty;$ see \cite{La:Lif} for more details. In
this paper we consider the simplest case when the magnetic field $H$ is a
constant and the electric force $F$ is a function of time $t$ (see Figure~1);
a more general case will be discussed elsewhere. Then the substitution%
\begin{equation}
\Psi\left(  \boldsymbol{r},t\right)  =e^{i\left(  xp_{x}+zp_{z}-S\left(
t,t_{0}\right)  \right)  /\hslash}\ \psi\left(  y,t\right)  , \label{ll5}%
\end{equation}
where%
\begin{equation}
S\left(  t,t_{0}\right)  =\left(  \frac{p_{z}^{2}}{2m}-\frac{\mu\sigma}%
{s}H\right)  \left(  t-t_{0}\right)  +\frac{cp_{x}}{eH}\int_{t_{0}}%
^{t}F\left(  \tau\right)  \ d\tau, \label{ll6}%
\end{equation}
results in the one-dimensional time-dependent Schr\"{o}dinger equation of the
harmonic oscillator driven by an external force in the $y$-direction
\begin{equation}
i\hslash\frac{\partial\psi}{\partial t}=-\frac{\hslash^{2}}{2m}\frac
{\partial^{2}\psi}{\partial y^{2}}+\frac{m\omega_{H}^{2}}{2}\left(
y-y_{0}\right)  ^{2}\psi-F\left(  t\right)  \left(  y-y_{0}\right)
\psi\label{ll7}%
\end{equation}
with%
\begin{equation}
\omega_{H}=\frac{\left\vert e\right\vert H}{mc},\qquad y_{0}=-\frac{cp_{x}%
}{eH}. \label{ll8}%
\end{equation}

The Cauchy initial value problem subject to special data%
\begin{equation}
\left.  \Psi\left(  \boldsymbol{r},t\right)  \right\vert _{t=t_{0}%
}=e^{i\left(  xp_{x}+zp_{z}\right)  /\hslash}\ \psi\left(  y,t_{0}\right)
=e^{i\left(  xp_{x}+zp_{z}\right)  /\hslash}\ \varphi\left(  y-y_{0}\right)
\label{ll9}%
\end{equation}
has the following solution%
\begin{equation}
\Psi\left(  \boldsymbol{r},t\right)  =\Psi\left(  \boldsymbol{r},t,p_{x}%
,p_{z}\right)  =e^{i\left(  xp_{x}+zp_{z}-S\left(  t,t_{0}\right)  \right)
/\hslash}\ \int_{-\infty}^{\infty}G\left(  y-y_{0},\eta,t,t_{0}\right)
\ \varphi\left(  \eta\right)  \ d\eta, \label{ll10}%
\end{equation}
where the propagator takes the form%
\begin{equation}
G\left(  y,\eta,t,t_{0}\right)  =G_{1}\left(  y,\eta,t-t_{0}\right)
\ e^{i\left(  a\left(  t,t_{0}\right)  y+b\left(  t,t_{0}\right)
\eta+c\left(  t,t_{0}\right)  \right)  } \label{ll11}%
\end{equation}
with%
\begin{align}
&  G_{1}\left(  y,\eta,t\right)  =\sqrt{\frac{m\omega_{H}}{2\pi i\hslash
\sin\omega_{H}t}}\ \label{ll12}\\
&  \quad\times\exp\left(  \dfrac{im\omega_{H}}{2\hslash\sin\omega_{H}t}\left(
\left(  y^{2}+\eta^{2}\right)  \cos\omega_{H}t-2y\eta\right)  \right)
,\nonumber
\end{align}%
\begin{align}
a\left(  t,t_{0}\right)   &  =\frac{1}{\hslash\sin\omega_{H}\left(
t-t_{0}\right)  }\ \int_{t_{0}}^{t}F\left(  \tau\right)  \sin\omega_{H}\left(
\tau-t_{0}\right)  \ d\tau,\label{ll13}\\
b\left(  t,t_{0}\right)   &  =-a\left(  t_{0},t\right)  \label{ll14}%
\end{align}
and%
\begin{equation}
c\left(  t,t_{0}\right)  =-\frac{\hslash}{2m}\int_{t_{0}}^{t}a^{2}\left(
\tau,t_{0}\right)  \ d\tau. \label{ll15}%
\end{equation}
See equations (\ref{spc7})--(\ref{spc12}) with $G\equiv0.$ Function $c\left(
t,t_{0}\right)  $ can be written in several different forms.

\subsection{Landau Levels}

In an absence of the external force $F\equiv0,$ equation (\ref{ll7}) is
formally identical to the time-dependent Schr\"{o}dinger equation for a simple
harmonic oscillator with the frequency $\omega_{H}.$ The standard substitution%
\begin{equation}
\psi\left(  y,t\right)  =e^{-i\varepsilon\left(  t-t_{0}\right)  /\hslash
}\ \chi\left(  y\right)  \label{ll16}%
\end{equation}
gives the corresponding stationary Schr\"{o}dinger equation as follows
\cite{La:Lif}%
\begin{equation}
\chi^{\prime\prime}+\frac{2m}{\hslash^{2}}\left(  \varepsilon-\frac{1}%
{2}m\omega_{H}^{2}\left(  y-y_{0}\right)  ^{2}\right)  \chi=0, \label{ll17}%
\end{equation}
which has the square integrable solutions only when%
\begin{equation}
\varepsilon=\hslash\omega_{H}\left(  n+\frac{1}{2}\right)  ,\qquad
n=0,1,2,...\ . \label{ll7a}%
\end{equation}
The eigenfunctions are%
\begin{equation}
\chi_{n}\left(  y\right)  =\frac{1}{\sqrt{2^{n}n!a_{H}\sqrt{\pi}}}%
\ \exp\left(  -\frac{\left(  y-y_{0}\right)  ^{2}}{2a_{H}^{2}}\right)
\ H_{n}\left(  \frac{y-y_{0}}{a_{H}}\right)  ,\quad a_{H}=\sqrt{\frac{\hslash
}{m\omega_{H}}}, \label{ll7b}%
\end{equation}
where $H_{n}\left(  \eta\right)  $ are the Hermite polynomials.\smallskip

Thus the total energy levels of a particle in a uniform magnetic field have
the form%
\begin{equation}
E_{n}=E_{n}\left(  p_{z},\sigma\right)  =\hslash\omega_{H}\left(  n+\frac
{1}{2}\right)  +\frac{p_{z}^{2}}{2m}-\frac{\mu\sigma}{s}H\qquad\left(
n=0,1,2,...\ \right)  . \label{ll18}%
\end{equation}
The first term here gives the discrete energy values corresponding to motion
in a plane perpendicular to the field. They are called Landau levels. The
expression (\ref{ll18}) does not contain the quantity $p_{x},$ which takes all
real values. Therefore the total energy levels are continuously degenerate.
For an electron, $\mu/s=-\left\vert e\right\vert \hslash/mc,$ and formula
(\ref{ll18}) becomes%
\begin{equation}
E_{n}=E_{n}\left(  p_{z},\sigma\right)  =\hslash\omega_{H}\left(  n+\frac
{1}{2}+\sigma\right)  +\frac{p_{z}^{2}}{2m}. \label{ll19}%
\end{equation}
In this case, there is an additional degeneracy: the levels with $n,$
$\sigma=1/2$ and $n+1,$ $\sigma=-1/2$ coincide: $E_{n}\left(  p_{z}%
,1/2\right)  =E_{n+1}\left(  p_{z},-1/2\right)  .$\smallskip

The three-dimensional wave functions corresponding to the energy levels
(\ref{ll18}) are given by%
\begin{equation}
\Psi_{n}\left(  \boldsymbol{r},t,\sigma\right)  =\Psi_{n}\left(
\boldsymbol{r},t,p_{x},p_{z},\sigma\right)  =e^{-iE_{n}\left(  p_{z}%
,\sigma\right)  \left(  t-t_{0}\right)  /\hslash}\ e^{i\left(  xp_{x}%
+zp_{z}\right)  /\hslash}\ \chi_{n}\left(  y\right)  . \label{ll20}%
\end{equation}
They are the eigenfunctions of the following set of commuting operators
$\widehat{p}_{x},$ $\widehat{p}_{z},$ $\widehat{s}_{z},$ and $\widehat{H}$
with $F\equiv0:$
\begin{equation}
\widehat{H}\Psi_{n}=E_{n}\Psi_{n},\quad\widehat{s}_{z}\Psi_{n}=\sigma\Psi
_{n},\quad\widehat{p}_{x}\Psi_{n}=p_{x}\Psi_{n},\quad\widehat{p}_{z}\Psi
_{n}=p_{z}\Psi_{n}. \label{ll19a}%
\end{equation}
The orthogonality relation in $\boldsymbol{R}^{3}$ is%
\begin{align}
&  \int_{\boldsymbol{R}^{3}}\Psi_{n}^{\ast}\left(  \boldsymbol{r}%
,t,p_{x},p_{z},\sigma\right)  \ \Psi_{m}\left(  \boldsymbol{r},t,p_{x}%
^{\prime},p_{z}^{\prime},\sigma^{\prime}\right)  \ dxdydz\label{ll20a}\\
&  \qquad=\left(  2\pi\hslash\right)  ^{2}\ \delta_{nm}\ \delta_{\sigma
\sigma^{\prime}}\ \delta\left(  p_{x}-p_{x}^{\prime}\right)  \delta\left(
p_{z}-p_{z}^{\prime}\right)  ,\nonumber
\end{align}
where%
\begin{equation}
\delta\left(  \alpha\right)  =\frac{1}{2\pi}\int_{-\infty}^{\infty}%
e^{i\alpha\xi}\ d\xi\label{ll20b}%
\end{equation}
is the Dirac delta function.

\subsection{Transition Amplitudes}

In the presence of external force, the quantum mechanical amplitude of a
transition between Landau's levels under the influence of the perpendicular
electric field can be explicitly found as a special case of our formulas
(\ref{effho5})--(\ref{effho7}). Indeed, solution (\ref{ll10}) takes the form%
\begin{equation}
\Psi\left(  \boldsymbol{r},t,\sigma\right)  =e^{-iS\left(  t,t_{0}\right)
/\hslash}\ \sum_{n=0}^{\infty}\Psi_{n}\left(  \boldsymbol{r},t_{0}%
,\sigma\right)  \sum_{m=0}^{\infty}c_{nm}\left(  t,t_{0}\right)
\ \int_{-\infty}^{\infty}\chi_{m}\left(  \eta\right)  \psi\left(  \eta
,t_{0}\right)  \ d\eta\label{ll21}%
\end{equation}
in view of the bilinear generating relation (\ref{effho4}). If $\psi\left(
y,t_{0}\right)  =\chi_{m}\left(  y\right)  ,$ this equation becomes%
\begin{equation}
\Psi\left(  \boldsymbol{r},t,\sigma\right)  =e^{-iS\left(  t,t_{0}\right)
/\hslash}\ \sum_{n=0}^{\infty}c_{nm}\left(  t,t_{0}\right)  \ \Psi_{n}\left(
\boldsymbol{r},t_{0},\sigma\right)  , \label{ll22}%
\end{equation}
where coefficients $c_{nm}\left(  t,t_{0}\right)  $ are given by
(\ref{effho6}) in terms of Charlier polynomials as follows%
\begin{align}
c_{nm}\left(  t,t_{0}\right)   &  =e^{i\left(  c-\left(  \omega_{H}\left(
t-t_{0}\right)  -ab\sin\omega_{H}\left(  t-t_{0}\right)  \right)  /2\right)
}\ e^{-\gamma^{2}/4}\label{ll23}\\
&  \quad\times\frac{i^{n+m}}{\sqrt{2^{n+m}n!m!}}\ \left(  a+b\delta\right)
^{n}\left(  b+a\delta\right)  ^{m}c_{m}^{\gamma^{2}/2}\left(  n\right)
\nonumber
\end{align}
with $\delta=e^{-i\omega_{H}\left(  t-t_{0}\right)  }$ and $\gamma^{2}%
=a^{2}+b^{2}+2ab\cos\omega_{H}\left(  t-t_{0}\right)  .$ Functions $a=a\left(
t,t_{0}\right)  ,$ $b=b\left(  t,t_{0}\right)  $ and $c=c\left(
t,t_{0}\right)  $ are evaluated by the integrals (\ref{ll13})--(\ref{ll15}),
respectively. The last two formulas (\ref{ll22})--(\ref{ll23}) and (\ref{ll6})
give us the quantum mechanical amplitude that the particle initially in Landau
state $m$ is found at time $t$ in state $n.$ For the particle initially in the
ground state $m=0,$ the probability to occupy state $n$ at time $t$ is given
by the Poisson distribution%
\begin{equation}
\left\vert c_{n0}\left(  t,t_{0}\right)  \right\vert ^{2}=e^{-\mu}\ \frac
{\mu^{n}}{n!},\qquad\mu=\frac{1}{2}\left(  a^{2}+b^{2}+2ab\cos\omega
_{H}\left(  t-t_{0}\right)  \right)  <1. \label{ll24}%
\end{equation}
The details are left to the reader.

\subsection{Propagator in Three Dimensions}

Our particular solutions (\ref{ll10}) subject to special initial data
(\ref{ll9}) have been constructed above as eigenfunctions of the operators
$\widehat{p}_{x}$ and $\widehat{p}_{z},$ whose continuous eigenvalues $p_{x}$
and $p_{z}$ vary from $-\infty$ to $\infty.$ By the superposition principle,
one can look for a general solution in $\boldsymbol{R}^{3}$ as a double
Fourier integral of the particular solution%
\begin{align}
\Psi\left(  \boldsymbol{r},t\right)   &  =%
%TCIMACRO{\diint _{-\infty}^{\infty}}%
%BeginExpansion
{\displaystyle\iint_{-\infty}^{\infty}}
%EndExpansion
a\left(  p_{x},p_{z}\right)  \Psi\left(  \boldsymbol{r},t,p_{x},p_{z}\right)
\ dp_{x}dp_{z}\label{ll25}\\
&  =%
%TCIMACRO{\diint _{-\infty}^{\infty}}%
%BeginExpansion
{\displaystyle\iint_{-\infty}^{\infty}}
%EndExpansion
dp_{x}dp_{z}\ a\left(  p_{x},p_{z}\right)  \ e^{i\left(  xp_{x}+zp_{z}\right)
/\hslash}\nonumber\\
&  \quad\times e^{-iS\left(  t,t_{0}\right)  /\hslash}\int_{-\infty}^{\infty
}G\left(  y-y_{0},\eta,t,t_{0}\right)  \ \varphi\left(  \eta\right)
\ d\eta,\nonumber
\end{align}
where functions $a\left(  p_{x},p_{z}\right)  $ do not depend on time $t$ and
$S\left(  t,t_{0}\right)  $ is given by (\ref{ll6}). Now we replace the
special initial data (\ref{ll9}) in $\boldsymbol{R}^{3}$ by the general one%
\begin{equation}
\left.  \Psi\left(  \boldsymbol{r},t\right)  \right\vert _{t=t_{0}}%
=\phi\left(  x,y,z\right)  , \label{ll26}%
\end{equation}
which is independent on $p_{x}$ (and $y_{0}$). Letting $t\rightarrow t_{0}$ in
(\ref{ll25}) and using the fundamental property of the Green function,%
\begin{equation}
\lim_{t\rightarrow t_{0}^{+}}\int_{-\infty}^{\infty}G\left(  y-y_{0}%
,\eta,t,t_{0}\right)  \ \varphi\left(  \eta\right)  \ d\eta=\varphi\left(
y-y_{0}\right)  , \label{ll26a}%
\end{equation}
one gets%
\begin{equation}
\phi\left(  x,y,z\right)  =%
%TCIMACRO{\diint _{-\infty}^{\infty}}%
%BeginExpansion
{\displaystyle\iint_{-\infty}^{\infty}}
%EndExpansion
a\left(  p_{x},p_{z}\right)  \ \varphi\left(  y-y_{0}\right)  \ e^{i\left(
xp_{x}+zp_{z}\right)  /\hslash}\ dp_{x}dp_{z}, \label{ll27}%
\end{equation}
where $y_{0}$ is a function of $p_{x}$ in view of (\ref{ll8}). Thus%
\begin{equation}
a\left(  p_{x},p_{z}\right)  \ \varphi\left(  y-y_{0}\right)  =\frac
{1}{\left(  2\pi\hslash\right)  ^{2}}%
%TCIMACRO{\diint _{-\infty}^{\infty}}%
%BeginExpansion
{\displaystyle\iint_{-\infty}^{\infty}}
%EndExpansion
\phi\left(  \xi,y,\zeta\right)  \ e^{-i\left(  \xi p_{x}+\zeta p_{z}\right)
/\hslash}\ d\xi d\zeta\label{ll28}%
\end{equation}
by the inverse of the Fourier transform. Its substitution into (\ref{ll25})
gives%
\begin{align}
\Psi\left(  \boldsymbol{r},t\right)   &  =\frac{1}{\left(  2\pi\hslash\right)
^{2}}%
%TCIMACRO{\diint _{-\infty}^{\infty}}%
%BeginExpansion
{\displaystyle\iint_{-\infty}^{\infty}}
%EndExpansion
dp_{x}dp_{z}\ e^{i\left(  xp_{x}+zp_{z}-S\left(  t,t_{0}\right)  \right)
/\hslash}\label{ll28a}\\
&  \quad\qquad\times\int_{-\infty}^{\infty}d\eta\ G\left(  y-y_{0},\eta
-y_{0},t,t_{0}\right) \nonumber\\
&  \qquad\qquad\times%
%TCIMACRO{\diint _{-\infty}^{\infty}}%
%BeginExpansion
{\displaystyle\iint_{-\infty}^{\infty}}
%EndExpansion
\phi\left(  \xi,\eta,\zeta\right)  \ e^{-i\left(  \xi p_{x}+\zeta
p_{z}\right)  /\hslash}\ d\xi d\zeta\nonumber
\end{align}
as a solution of our initial value problem. A familiar integral form of this
solution is as follows%
\begin{equation}
\Psi\left(  \boldsymbol{r},t\right)  =\int_{\boldsymbol{R}^{3}}G\left(
\boldsymbol{r},\boldsymbol{\rho},t,t_{0}\right)  \ \phi\left(  \xi,\eta
,\zeta\right)  \ d\xi d\eta d\zeta, \label{ll29}%
\end{equation}
where the Green function (propagator) is given as a double Fourier integral%
\begin{align}
G\left(  \boldsymbol{r},\boldsymbol{\rho},t,t_{0}\right)   &  =\frac
{1}{\left(  2\pi\hslash\right)  ^{2}}%
%TCIMACRO{\diint _{-\infty}^{\infty}}%
%BeginExpansion
{\displaystyle\iint_{-\infty}^{\infty}}
%EndExpansion
e^{i\left(  \left(  x-\xi\right)  p_{x}+\left(  z-\zeta\right)  p_{z}\right)
/\hslash}\ e^{-iS\left(  t,t_{0}\right)  /\hslash}\label{ll30}\\
&  \qquad\qquad\quad\times G\left(  y-y_{0},\eta-y_{0},t,t_{0}\right)
\ dp_{x}dp_{z}\nonumber
\end{align}
with the help of the Fubini theorem.\smallskip

This integral can be evaluated in terms of elementary functions as follows.
Integration over $p_{z}$ gives the free particle propagator of a motion in the
direction of magnetic field%
\begin{align}
G_{0}\left(  z-\zeta,t-t_{0}\right)   &  =\frac{1}{2\pi\hslash}%
%TCIMACRO{\dint _{-\infty}^{\infty}}%
%BeginExpansion
{\displaystyle\int_{-\infty}^{\infty}}
%EndExpansion
\exp\left(  \dfrac{i}{\hslash}\left(  \left(  z-\zeta\right)  p_{z}%
-\dfrac{p_{z}^{2}}{2m}\left(  t-t_{0}\right)  \right)  \right)  \ dp_{z}%
\label{ll31}\\
&  =\sqrt{\frac{m}{2\pi i\hslash\left(  t-t_{0}\right)  }}\ \exp\left(
\frac{im\left(  z-\zeta\right)  ^{2}}{2\hslash\left(  t-t_{0}\right)  }\right)
\nonumber
\end{align}
by the integral (\ref{green6}). Thus%
\begin{align}
&  G\left(  \boldsymbol{r},\boldsymbol{\rho},t,t_{0}\right)  =\exp\left(
\dfrac{i\mu\sigma H}{\hslash s}\left(  t-t_{0}\right)  \right)  \ G_{0}\left(
z-\zeta,t-t_{0}\right) \label{ll32}\\
&  \quad\times\frac{1}{2\pi\hslash}\int_{-\infty}^{\infty}\exp\left(
\dfrac{i}{\hslash}\left(  x-\xi\right)  p_{x}\right)  \ \exp\left(
-\dfrac{icp_{x}}{\hslash eH}%
%TCIMACRO{\dint _{t_{0}}^{t}}%
%BeginExpansion
{\displaystyle\int_{t_{0}}^{t}}
%EndExpansion
F\left(  \tau\right)  \ d\tau\right) \nonumber\\
&  \quad\qquad\qquad\times G\left(  y-y_{0},\eta-y_{0},t,t_{0}\right)
\ dp_{x}\nonumber\\
&  \qquad=\exp\left(  \dfrac{i\mu\sigma H}{\hslash s}\left(  t-t_{0}\right)
\right)  \ G_{0}\left(  z-\zeta,t-t_{0}\right) \nonumber\\
&  \quad\quad\quad\times G_{1}\left(  y,\eta,t-t_{0}\right)  \ e^{i\left(
a\left(  t,t_{0}\right)  y+b\left(  t,t_{0}\right)  \eta+c\left(
t,t_{0}\right)  \right)  }\nonumber\\
&  \quad\quad\quad\times\frac{1}{2\pi\hslash}\int_{-\infty}^{\infty}%
\exp\left(  \dfrac{ip_{x}}{\hslash}\left(  x-\xi-\dfrac{c}{eH}%
%TCIMACRO{\dint _{t_{0}}^{t}}%
%BeginExpansion
{\displaystyle\int_{t_{0}}^{t}}
%EndExpansion
F\left(  \tau\right)  \ d\tau\right)  \right) \nonumber\\
&  \quad\quad\quad\quad\times\exp\left(  \dfrac{i\left(  a\left(
t,t_{0}\right)  +b\left(  t,t_{0}\right)  \right)  cp_{x}}{eH}\right)
\nonumber\\
&  \quad\quad\quad\quad\times\exp\left(  \dfrac{-i}{\hslash}\left(
\frac{p_{x}^{2}}{m\omega_{H}}+\frac{\left\vert e\right\vert }{e}\left(
y+\eta\right)  p_{x}\right)  \tan\left(  \omega_{H}\left(  t-t_{0}\right)
/2\right)  \right)  \ dp_{x}.\nonumber
\end{align}
In view of (\ref{green6}), the last integral is given by%
\begin{align}
&  \frac{1}{2\pi\hslash}\int_{-\infty}^{\infty}\exp\left(  \dfrac{ip_{x}%
}{\hslash}\left(  x-\xi-\dfrac{c}{eH}%
%TCIMACRO{\dint _{t_{0}}^{t}}%
%BeginExpansion
{\displaystyle\int_{t_{0}}^{t}}
%EndExpansion
F\left(  \tau\right)  \ d\tau\right)  \right) \label{ll33}\\
&  \quad\qquad\times\exp\left(  \dfrac{i\left(  a\left(  t,t_{0}\right)
+b\left(  t,t_{0}\right)  \right)  cp_{x}}{eH}\right) \nonumber\\
&  \quad\qquad\times\exp\left(  \dfrac{-i}{\hslash}\left(  \frac{p_{x}^{2}%
}{m\omega_{H}}+\frac{e}{\left\vert e\right\vert }\left(  y+\eta\right)
p_{x}\right)  \tan\left(  \omega_{H}\left(  t-t_{0}\right)  /2\right)
\right)  \ dp_{x}\nonumber\\
&  \qquad=\sqrt{\frac{m\omega_{H}\cot\left(  \omega_{H}\left(  t-t_{0}\right)
/2\right)  }{4\pi i\hslash}}\ \exp\left(  \dfrac{im\omega_{H}\cot\left(
\omega_{H}\left(  t-t_{0}\right)  /2\right)  }{4\hslash}\ \beta^{2}\right)
,\nonumber
\end{align}
where%
\begin{equation}
\beta=x-\xi-\frac{e}{\left\vert e\right\vert }\left(  y+\eta\right)
\tan\left(  \omega_{H}\left(  t-t_{0}\right)  /2\right)  +d\left(
t,t_{0}\right)  \label{ll33a}%
\end{equation}
with%
\begin{align}
&  d\left(  t,t_{0}\right)  =\frac{c}{eH\sin\omega_{H}\left(  t-t_{0}\right)
}\label{ll34}\\
&  \quad\quad\quad\times%
%TCIMACRO{\dint _{t_{0}}^{t}}%
%BeginExpansion
{\displaystyle\int_{t_{0}}^{t}}
%EndExpansion
F\left(  \tau\right)  \left(  \sin\omega_{H}\left(  \tau-t_{0}\right)
-\sin\omega_{H}\left(  \tau-t\right)  -\sin\omega_{H}\left(  t-t_{0}\right)
\right)  \ d\tau.\nonumber
\end{align}
Here we have used (\ref{ll13})--(\ref{ll14}). As a result, we arrive at the
following factorization of our propagator%
\begin{align}
G\left(  \boldsymbol{r},\boldsymbol{\rho},t,t_{0}\right)   &  =G_{0}\left(
z-\zeta,t-t_{0}\right)  \ G_{1}\left(  y,\eta,t-t_{0}\right)  \ e^{i\left(
a\left(  t,t_{0}\right)  y+b\left(  t,t_{0}\right)  \eta+c\left(
t,t_{0}\right)  \right)  }\label{ll35}\\
&  \quad\times G_{2}\left(  x,\xi,y,\eta,t,t_{0}\right)  ,\nonumber
\end{align}
where $G_{0}\left(  z-\zeta,t-t_{0}\right)  $ is the free particle propagator
in (\ref{ll31}), $G_{1}\left(  y,\eta,t-t_{0}\right)  $ is the simple harmonic
oscillator propagator in (\ref{ll12}), and
\begin{align}
&  G_{2}\left(  x,\xi,y,\eta,t,t_{0}\right)  =\exp\left(  \dfrac{i\mu\sigma
H}{\hslash s}\left(  t-t_{0}\right)  \right) \label{ll35a}\\
&  \quad\times\sqrt{\frac{m\omega_{H}\cot\left(  \omega_{H}\left(
t-t_{0}\right)  /2\right)  }{4\pi i\hslash}}\ \exp\left(  \dfrac{im\omega
_{H}\cot\left(  \omega_{H}\left(  t-t_{0}\right)  /2\right)  }{4\hslash
}\ \beta^{2}\right) \nonumber
\end{align}
with $\beta=\beta\left(  x,\xi,y,\eta,t,t_{0}\right)  $ given by
(\ref{ll33a})--(\ref{ll34}).\smallskip

Our propagator can be simplified to a somewhat more convenient form as follows%
\begin{align}
G\left(  \boldsymbol{r},\boldsymbol{\rho},t,t_{0}\right)   &  =G_{0}\left(
z-\zeta,t-t_{0}\right)  \ G_{H}\left(  x,\xi,y,\eta,t-t_{0}\right)
\label{ll35b}\\
&  \quad\times G_{F}\left(  x,\xi,y,\eta,t,t_{0}\right)  .\nonumber
\end{align}
Here $G_{0}\left(  z,t\right)  $ is the free particle propagator in the
direction of magnetic field. The function%
\begin{align}
&  G_{H}\left(  x,\xi,y,\eta,t\right)  =\exp\left(  \dfrac{i\mu\sigma
Ht}{\hslash s}\right)  \ \frac{m\omega_{H}}{4\pi i\hslash\sin\left(
\omega_{H}t/2\right)  }\ \label{ll35c}\\
&  \quad\times\exp\left(  \dfrac{im\omega_{H}}{4\hslash}\left(  \left(
\left(  x-\xi\right)  ^{2}+\left(  y-\eta\right)  ^{2}\right)  \cot\left(
\omega_{H}t/2\right)  -2\frac{e}{\left\vert e\right\vert }\left(
x-\xi\right)  \left(  y+\eta\right)  \right)  \right) \nonumber
\end{align}
is the propagator corresponding to a motion in a plane perpendicular to the
magnetic field in an absence of the electric field (compare our expression
with one in \cite{Fey:Hib}, where $F=\mu=0,$ and see a remark below in order
to establish an identity of two results). The third factor%
\begin{equation}
G_{F}\left(  x,\xi,y,\eta,t,t_{0}\right)  =e^{iW_{F}\left(  t,t_{0}\right)
/\hslash} \label{ll35d}%
\end{equation}
with%
\begin{align}
&  W_{F}\left(  t,t_{0}\right)  =\hslash\left(  a\left(  t,t_{0}\right)
y+b\left(  t,t_{0}\right)  \eta+c\left(  t,t_{0}\right)  \right)
\label{ll35e}\\
&  \quad+\frac{1}{4}m\omega_{H}\ d\left(  t,t_{0}\right)  \left(  \left(
d\left(  t,t_{0}\right)  +2\left(  x-\xi\right)  \right)  \cot\left(
\omega_{H}\left(  t-t_{0}\right)  /2\right)  -2\frac{e}{\left\vert
e\right\vert }\left(  y+\eta\right)  \right) \nonumber
\end{align}
is a contribution from the electric field. When $F=0,$ $W_{F}=0$ and
$G_{F}=1.$\smallskip

The solution of the Cauchy initial value problem in $\boldsymbol{R}^{3}$
subject to the general initial data%
\begin{equation}
\left.  \Psi\left(  \boldsymbol{r},t\right)  \right\vert _{t=t_{0}}%
=\Psi\left(  \boldsymbol{r},t_{0}\right)  =\phi\left(  x,y,z\right)
\label{ll36}%
\end{equation}
has the form%
\begin{equation}
\Psi\left(  \boldsymbol{r},t\right)  =\int_{\boldsymbol{R}^{3}}G\left(
\boldsymbol{r},\boldsymbol{\rho},t,t_{0}\right)  \ \Psi\left(
\boldsymbol{\rho},t_{0}\right)  \ d\xi d\eta d\zeta, \label{ll37}%
\end{equation}
which gives explicitly the time evolution operator for a motion of a charged
particle in uniform perpendicular magnetic and electric fields with a given
projection of the spin $s_{z}=\sigma$ in the direction of magnetic field. By
choosing $\Psi\left(  \boldsymbol{r},t_{0}\right)  =\delta\left(
\boldsymbol{r}-\boldsymbol{r}_{0}\right)  ,$ where $\delta\left(
\boldsymbol{r}\right)  $ is the Dirac delta function in three dimensions, we
formally obtain%
\begin{equation}
\Psi\left(  \boldsymbol{r},t\right)  =G\left(  \boldsymbol{r},\boldsymbol{r}%
_{0},t,t_{0}\right)  \label{ll38}%
\end{equation}
as the wave function at time $t$ of the particle initially located at a point
$\boldsymbol{r}=\boldsymbol{r}_{0}.$ Then equation (\ref{ll37}) gives a
general solution by the superposition principle.\smallskip

\noindent\textbf{Remark.\/} The vector potential of the uniform magnetic field
in the $z$-direction is defined up to a gauge transformation \cite{La:Lif}%
\begin{equation}
\boldsymbol{A}=-yH\ \boldsymbol{e}_{x}\rightarrow\boldsymbol{A}^{\prime
}=\boldsymbol{A}+\nabla f=-\frac{1}{2}yH\ \boldsymbol{e}_{x}+\frac{1}%
{2}xH\ \boldsymbol{e}_{y}=\frac{1}{2}\boldsymbol{H\times r}\label{ll39}%
\end{equation}
with $f\left(  x,y\right)  =xyH/2.$ The corresponding transformation of the
wave function is given by%
\begin{equation}
\Psi\rightarrow\Psi^{\prime}=\Psi\exp\left(  \frac{ief\left(  x,y\right)
}{\hslash c}\right)  =\Psi\exp\left(  \dfrac{im\omega_{H}}{4\hslash}\left(
2\frac{e}{\left\vert e\right\vert }xy\right)  \right)  \label{ll40}%
\end{equation}
and in view of (\ref{ll37})%
\begin{align}
G_{H} &  \rightarrow\exp\left(  \frac{ief\left(  x,y\right)  }{\hslash
c}\right)  G_{H}\exp\left(  -\frac{ief\left(  \xi,\eta\right)  }{\hslash
c}\right)  =\frac{m\omega_{H}}{4\pi i\hslash\sin\left(  \omega_{H}t/2\right)
}\label{ll41}\\
&  \times\exp\left(  \dfrac{im\omega_{H}}{4\hslash}\left(  \left(  \left(
x-\xi\right)  ^{2}+\left(  y-\eta\right)  ^{2}\right)  \cot\left(  \omega
_{H}t/2\right)  -2\frac{e}{\left\vert e\right\vert }\left(  x\eta-\xi
y\right)  \right)  \right)  ,\nonumber
\end{align}
which is, essentially, equation (3-64) on page~64 of \cite{Fey:Hib}, where we
have corrected a typo. The details are left to the reader.

\section{Diffusion-Type Equation}

\subsection{Special Case}

A formal substitution of $t\rightarrow-it$ and $\psi\rightarrow u$ into
equation (\ref{solu1}) with $\omega=2\kappa$ and $\sqrt{2\mu}=\varepsilon$
yields the following time-dependent diffusion-type equation%
\begin{equation}
\frac{\partial u}{\partial t}=\kappa\left(  \frac{\partial^{2}u}{\partial
x^{2}}-x^{2}u\right)  -\varepsilon\left(  \left(  \cosh\left(  2\kappa
-1\right)  t\right)  \ xu+\left(  \sinh\left(  2\kappa-1\right)  t\right)
\ \frac{\partial u}{\partial x}\right)  , \label{diff1}%
\end{equation}
where the initial condition is%
\begin{equation}
\left.  u\left(  x,t\right)  \right\vert _{t=0}=u_{0}\left(  x\right)
\qquad\left(  -\infty<x<\infty\right)  . \label{diff2}%
\end{equation}
As in the case of the time-dependent Schr\"{o}dinger equation, in order to
solve this initial value problem, we use the eigenfunction expansion method.
Hence the solution is given by%
\begin{equation}
u\left(  x,t\right)  =\sum_{n=0}^{\infty}\Psi_{n}\left(  x\right)  \sum
_{m=0}^{\infty}c_{nm}\left(  t\right)  \ \int_{-\infty}^{\infty}\Psi
_{m}\left(  y\right)  u_{0}\left(  y\right)  \ dy, \label{diff3}%
\end{equation}
where%
\begin{align}
c_{nm}\left(  t\right)   &  =\left(  -1\right)  ^{n-m}\frac{\varepsilon^{n+m}%
}{\sqrt{2^{n+m}n!m!}}\ e^{-\left(  \varepsilon^{2}/2\right)  \left(
1-e^{-t}\right)  }\ e^{-\left(  \left(  2\kappa-1\right)  n+\kappa
-\varepsilon^{2}/2\right)  t}\label{diff4}\\
&  \quad\times\left(  1-e^{-t}\right)  ^{m+n}\ _{2}F_{0}\left(
-n,\ -m;\ \frac{2e^{-t}}{\varepsilon^{2}\left(  1-e^{-t}\right)  ^{2}}\right)
\nonumber
\end{align}
by analytic continuation $t\rightarrow-it$ with $\omega=2\kappa$ and
$\mu=\varepsilon^{2}/2<1$\ in (\ref{solu4}) and (\ref{solu7}). One can easily
verify that%
\[
\lim_{t\rightarrow0^{+}}c_{nm}\left(  t\right)  =\delta_{nm}%
\]
and that if $0<\varepsilon<\sqrt{2}$ and $\kappa\geq1/2,$ $\kappa
>\varepsilon^{2}/2,$%
\[
\lim_{t\rightarrow\infty}c_{nm}\left(  t\right)  =0\qquad\left(
m,n=0,1,2,...\ \right)  .
\]
Thus the limiting distribution is%
\begin{equation}
\lim_{t\rightarrow\infty}u\left(  x,t\right)  \equiv0\qquad\left(
-\infty<x<\infty\right)  , \label{diff5}%
\end{equation}
which is independent of the initial data (\ref{diff2}).\smallskip

Relation (\ref{heis8}) becomes%
\begin{align}
u\left(  x,t\right)   &  =e^{-\left(  \varepsilon^{2}/2\right)  \sinh
t-\left(  \kappa-\varepsilon^{2}/2\right)  t}\sum_{n=0}^{\infty}\left(
-1\right)  ^{n}e^{-t\left(  2\kappa-1/2\right)  n}\ \Psi_{n}\left(  x\right)
\label{diff6}\\
&  \quad\times\sum_{m=0}^{\infty}t_{mn}\left(  \beta\right)  \int_{-\infty
}^{\infty}\left(  -1\right)  ^{m}e^{-mt/2}\ \Psi_{m}\left(  y\right)
u_{0}\left(  y\right)  \ dy\nonumber
\end{align}
with $\beta=\beta\left(  t\right)  =-2i\varepsilon\sinh\left(  t/2\right)  .$
With the help of (\ref{four1a}), (\ref{heis6}) and the Fubuni theorem we
transform%
\begin{align}
&  \sum_{m=0}^{\infty}t_{mn}\left(  \beta\right)  \int_{-\infty}^{\infty
}e^{-mt/2}\ \Psi_{m}\left(  -y\right)  u_{0}\left(  y\right)
\ dy\label{diff6a}\\
&  \quad=\int\int_{-\infty}^{\infty}K_{e^{-t/2}}\left(  -y,z\right)  \left(
\sum_{m=0}^{\infty}t_{mn}\left(  \beta\right)  \Psi_{m}\left(  z\right)
\right)  u_{0}\left(  y\right)  \ dydz\nonumber\\
&  \quad=\int\int_{-\infty}^{\infty}K_{e^{-t/2}}\left(  -y,z\right)  \left(
e^{\gamma z}\Psi_{n}\left(  z\right)  \right)  u_{0}\left(  y\right)
\ dydz,\nonumber
\end{align}
where $\gamma=i\beta=2\varepsilon\sinh\left(  t/2\right)  .$ The series
(\ref{diff6}) becomes%
\begin{align}
u\left(  x,t\right)   &  =e^{-\left(  \varepsilon^{2}/2\right)  \sinh
t-\left(  \kappa-\varepsilon^{2}/2\right)  t}\sum_{n=0}^{\infty}e^{-t\left(
2\kappa-1/2\right)  n}\ \Psi_{n}\left(  -x\right) \label{diff6b}\\
&  \quad\times\int\int_{-\infty}^{\infty}K_{e^{-t/2}}\left(  -y,z\right)
\left(  e^{\gamma z}\Psi_{n}\left(  z\right)  \right)  u_{0}\left(  y\right)
\ dydz\nonumber\\
&  =e^{-\left(  \varepsilon^{2}/2\right)  \sinh t-\left(  \kappa
-\varepsilon^{2}/2\right)  t}\int\int_{-\infty}^{\infty}K_{e^{-t/2}}\left(
-y,z\right)  e^{\gamma z}\nonumber\\
&  \quad\times\left(  \sum_{n=0}^{\infty}e^{-t\left(  2\kappa-1/2\right)
n}\ \Psi_{n}\left(  -x\right)  \Psi_{n}\left(  z\right)  \right)  u_{0}\left(
y\right)  \ dydz\nonumber\\
&  =e^{-\left(  \varepsilon^{2}/2\right)  \sinh t-\left(  \kappa
-\varepsilon^{2}/2\right)  t}\nonumber\\
&  \quad\times\int_{-\infty}^{\infty}\left(  \int_{-\infty}^{\infty
}K_{e^{-t/2}}\left(  -y,z\right)  e^{\gamma z}K_{e^{-t\left(  2\kappa
-1/2\right)  }}\left(  -x,z\right)  \ dz\right)  u_{0}\left(  y\right)
\ dy\nonumber
\end{align}
in view of the generating relation (\ref{four1}). Therefore, the integral form
of the solution (\ref{diff3})--(\ref{diff4}) is%
\begin{equation}
u\left(  x,t\right)  =e^{-\left(  \varepsilon^{2}/2\right)  \sinh t-\left(
\kappa-\varepsilon^{2}/2\right)  t}\int_{-\infty}^{\infty}\mathcal{H}%
_{t}\left(  x,y\right)  u_{0}\left(  y\right)  \ dy, \label{diff7}%
\end{equation}
where by the definition%
\begin{equation}
\mathcal{H}_{t}\left(  x,y\right)  :=\int_{-\infty}^{\infty}K_{e^{-t\left(
2\kappa-1/2\right)  }}\left(  -x,z\right)  e^{\gamma z}K_{e^{-t/2}}\left(
-y,z\right)  \ dz. \label{diff8}%
\end{equation}
Denoting $r_{1}=e^{-t\left(  2\kappa-1/2\right)  }$ and $r_{2}=e^{-t/2}$ we
obtain by (\ref{four1}) that
\begin{align}
&  \int_{-\infty}^{\infty}K_{r_{1}}\left(  -x,z\right)  e^{\gamma z}K_{r_{2}%
}\left(  -y,z\right)  \ dz\label{diff9}\\
&  \quad=\frac{1}{\pi\sqrt{\left(  1-r_{1}^{2}\right)  \left(  1-r_{2}%
^{2}\right)  }}\ \exp\left(  -\frac{\left(  1+r_{1}^{2}\right)  \left(
1-r_{2}^{2}\right)  x^{2}+\left(  1-r_{1}^{2}\right)  \left(  1+r_{2}%
^{2}\right)  y^{2}}{2\left(  1-r_{1}^{2}\right)  \left(  1-r_{2}^{2}\right)
}\right) \nonumber\\
&  \qquad\times\int_{-\infty}^{\infty}\exp\left(  \frac{\left(  1-r_{1}%
^{2}\right)  \left(  1-r_{2}^{2}\right)  \gamma-2r_{1}\left(  1-r_{2}%
^{2}\right)  x-2r_{2}\left(  1-r_{1}^{2}\right)  y}{\left(  1-r_{1}%
^{2}\right)  \left(  1-r_{2}^{2}\right)  }\ z\right) \nonumber\\
&  \qquad\qquad\qquad\times\exp\left(  -\frac{1-r_{1}^{2}r_{2}^{2}}{\left(
1-r_{1}^{2}\right)  \left(  1-r_{2}^{2}\right)  }\ z^{2}\right)  \ dz\nonumber
\end{align}
and the integral can be evaluated with the help of an elementary formula%
\begin{equation}
\int_{-\infty}^{\infty}e^{-az^{2}+2bz}\ dz=\sqrt{\frac{\pi}{a}}\ e^{b^{2}%
/a},\qquad a>0. \label{diff10}%
\end{equation}
As a result, an analog of the heat kernel in (\ref{diff7})--(\ref{diff8}) is
given by%
\begin{align}
&  \mathcal{H}_{t}\left(  x,y\right)  =\frac{1}{\sqrt{\pi\left(  1-r_{1}%
^{2}r_{2}^{2}\right)  }}\ \exp\left(  -\frac{\left(  1-r_{1}^{2}r_{2}%
^{2}\right)  \left(  x^{2}+y^{2}\right)  +\left(  r_{1}^{2}-r_{2}^{2}\right)
\left(  x^{2}-y^{2}\right)  }{2\left(  1-r_{1}^{2}\right)  \left(  1-r_{2}%
^{2}\right)  }\right) \label{diff11}\\
&  \times\exp\left(  \frac{\left[  \left(  1-r_{1}^{2}\right)  \left(
1-r_{2}^{2}\right)  \gamma-\left(  r_{1}+r_{2}\right)  \left(  1-r_{1}%
r_{2}\right)  \left(  x+y\right)  -\left(  r_{1}-r_{2}\right)  \left(
1+r_{1}r_{2}\right)  \left(  x-y\right)  \right]  ^{2}}{4\left(  1-r_{1}%
^{2}\right)  \left(  1-r_{2}^{2}\right)  \left(  1-r_{1}^{2}r_{2}^{2}\right)
}\right)  \ \nonumber
\end{align}
with $r_{1}=e^{-t\left(  2\kappa-1/2\right)  },$ $r_{2}=e^{-t/2}$ and
$\gamma=2\varepsilon\sinh\left(  t/2\right)  ,$ $t>0.$ The last expression can
be simplified to a somewhat more convenient form%
\begin{align}
&  \mathcal{H}_{t}\left(  x,y\right)  =\exp\left(  -\left(  \frac{r_{1}+r_{2}%
}{1+r_{1}r_{2}}\left(  x+y\right)  +\frac{r_{1}-r_{2}}{1-r_{1}r_{2}}\left(
x-y\right)  \right)  \ \frac{\gamma}{2}\right) \label{diff11a}\\
&  \qquad\qquad\quad\times\exp\left(  \frac{\left(  1-r_{1}^{2}\right)
\left(  1-r_{2}^{2}\right)  }{1-r_{1}^{2}r_{2}^{2}}\ \frac{\gamma^{2}}%
{4}\right)  \ K_{r_{1}r_{2}}\left(  x,y\right) \nonumber
\end{align}
in terms of the Mehler kernel (\ref{four1}). One can show that%
\begin{equation}
\lim_{t\rightarrow0^{+}}u\left(  x,t\right)  =u_{0}\left(  x\right)
\label{diff11b}%
\end{equation}
by methods of \cite{Wi}. The details are left to the reader.\smallskip

A formal substitution of $u_{0}\left(  x\right)  =\delta\left(  x-x_{0}%
\right)  $ into (\ref{diff7}) gives%
\begin{equation}
u\left(  x,t\right)  =H\left(  x,x_{0},t\right)  =e^{-\left(  \varepsilon
^{2}/2\right)  \sinh t-\left(  \kappa-\varepsilon^{2}/2\right)  t}%
\ \mathcal{H}_{t}\left(  x,x_{0}\right)  \label{diff12}%
\end{equation}
as the fundamental solution of the diffusion equation (\ref{diff1}).\smallskip

In the limit $\varepsilon\rightarrow0$ we obtain%
\begin{equation}
u\left(  x,t\right)  =e^{-\kappa t}\ \int_{-\infty}^{\infty}K_{e^{-2\kappa t}%
}\left(  x,y\right)  \ u_{0}\left(  y\right)  \ dy \label{diff12a}%
\end{equation}
as the exact solution of the corresponding initial value problem in terms of
the Mehler kernel (\ref{four1}). This kernel gives also a familiar expression
in statistical mechanics for the density matrix for a system consisting of a
simple harmonic oscillator \cite{Fey:Hib}.

\subsection{Generalization}

A formal substitution of $t\rightarrow-it$ and $\psi\rightarrow u$ into
(\ref{fho1}) with $\omega=2\kappa$ and $f\rightarrow f,$ $g\rightarrow-ig$
yields a diffusion-type equation%
\begin{equation}
\frac{\partial u}{\partial t}=\kappa\left(  \frac{\partial^{2}u}{\partial
x^{2}}-x^{2}u\right)  +f\left(  t\right)  \ xu-g\left(  t\right)
\ \frac{\partial u}{\partial x}, \label{diff13}%
\end{equation}
where $f\left(  t\right)  $ and $g\left(  t\right)  $ are real valued
functions of time, subject to the initial condition%
\begin{equation}
\left.  u\left(  x,t\right)  \right\vert _{t=0}=u_{0}\left(  x\right)
\qquad\left(  -\infty<x<\infty\right)  . \label{diff14}%
\end{equation}
The exact solution is%
\begin{equation}
u\left(  x,t\right)  =\int_{-\infty}^{\infty}H\left(  x,y,t\right)
u_{0}\left(  y\right)  \ dy \label{diff15}%
\end{equation}
and the Green function can be found in the form%
\begin{equation}
H\left(  x,y,t\right)  =H_{0}\left(  x,y,t\right)  \ e^{a\left(  t\right)
x+b\left(  t\right)  y+c\left(  t\right)  }, \label{diff16}%
\end{equation}
where%
\begin{equation}
H_{0}\left(  x,y,t\right)  =\sqrt{\frac{r}{\pi\left(  1-r^{2}\right)  }}%
\ \exp\left(  \frac{4xyr-\left(  x^{2}+y^{2}\right)  \left(  1+r^{2}\right)
}{2\left(  1-r^{2}\right)  }\right)  \label{diff17}%
\end{equation}
with $r=e^{-2\kappa t}.$ Indeed, substitution of (\ref{diff16}) into
(\ref{diff13}) gives the system of equations%
\begin{align}
\frac{d}{dt}\left(  \sinh\left(  2\kappa t\right)  \ a\left(  t\right)
\right)   &  =f\left(  t\right)  \sinh\left(  2\kappa t\right)  +g\left(
t\right)  \cosh\left(  2\kappa t\right)  ,\label{diff18a}\\
\frac{d}{dt}b\left(  t\right)   &  =\frac{2\kappa a\left(  t\right)  -g\left(
t\right)  }{\sinh\left(  2\kappa t\right)  },\label{diff19a}\\
\frac{d}{dt}c\left(  t\right)   &  =\kappa a^{2}\left(  t\right)  -g\left(
t\right)  a\left(  t\right)  \label{diff20a}%
\end{align}
and the solutions are%
\begin{align}
a\left(  t\right)   &  =\frac{1}{\sinh\left(  2\kappa t\right)  }\int_{0}%
^{t}\left(  f\left(  s\right)  \sinh\left(  2\kappa s\right)  +g\left(
s\right)  \cosh\left(  2\kappa s\right)  \right)  \ ds,\label{diff18}\\
b\left(  t\right)   &  =\int_{0}^{t}\frac{2\kappa a\left(  s\right)  -g\left(
s\right)  }{\sinh\left(  2\kappa s\right)  }\ ds,\label{diff19}\\
c\left(  t\right)   &  =\int_{0}^{t}\left(  \kappa a^{2}\left(  s\right)
-g\left(  s\right)  a\left(  s\right)  \right)  \ ds \label{diff20}%
\end{align}
provided $a\left(  0\right)  =b\left(  0\right)  =c\left(  0\right)
=0.$\smallskip

An analog of the expansion (\ref{effho4}) is%
\begin{equation}
H\left(  x,y,t\right)  =\sum_{n=0}^{\infty}\sum_{m=0}^{\infty}c_{nm}\left(
t\right)  \ \Psi_{n}\left(  x\right)  \Psi_{m}\left(  y\right)  \label{diff21}%
\end{equation}
with%
\begin{align}
c_{nm}\left(  t\right)   &  =\frac{1}{\sqrt{2^{n+m}n!m!}}\ e^{c-\kappa
t-\left(  ab/2\right)  \sinh\left(  2\kappa t\right)  +\lambda^{2}%
/4}\label{diff22}\\
&  \qquad\times\left(  a+br\right)  ^{n}\left(  b+ar\right)  ^{m}\ _{2}%
F_{0}\left(  -n,\ -m;\ \frac{2}{\lambda^{2}}\right)  .\nonumber
\end{align}
Here $r=e^{-2\kappa t},$ $\lambda^{2}=a^{2}+b^{2}+2ab\cosh\left(  2\kappa
t\right)  $ and functions $a\left(  t\right)  ,$ $b\left(  t\right)  $ and
$c\left(  t\right)  $ are given by the integrals (\ref{diff18})--(\ref{diff20}%
), respectively. This can be derived by expanding the kernel (\ref{diff16}) in
the double series in the same fashion as in section~11, or by the substitution
$t\rightarrow-it,$ $a\rightarrow-ia,$ $b\rightarrow-ib,$ and $c\rightarrow-ic$
in (\ref{effho4}). The coefficients $c_{nm}\left(  t\right)  $ are positive
when $t>0.$\smallskip

The solution (\ref{diff15}) takes the form%
\begin{equation}
u\left(  x,t\right)  =\sum_{n=0}^{\infty}\Psi_{n}\left(  x\right)  \sum
_{m=0}^{\infty}c_{nm}\left(  t\right)  \ \int_{-\infty}^{\infty}\Psi
_{m}\left(  y\right)  u_{0}\left(  y\right)  \ dy. \label{diff23}%
\end{equation}
These results can be extended to the case when parameter $\kappa$ is a
function of time in equation (\ref{diff13}). The details are left to the
reader.\smallskip

\noindent\textbf{Acknowledgment.\/} This paper is written as a part of the
summer 2007 program on analysis of the Mathematical and Theoretical Biology
Institute (MTBI) at Arizona State University. The MTBI/SUMS undergraduate
research program is supported by The National Science Foundation
(DMS--0502349), The National Security Agency (DOD--H982300710096), The Sloan
Foundation, and Arizona State University. The authors are grateful to
Professor Carlos Castillo-Ch\'{a}vez for support and reference
\cite{Bet:Cin:Kai:Cas}. We thank Professors George Andrews, George Gasper,
Slim Ibrahim, Hunk Kuiper, Mizan Rahman, Svetlana Roudenko, and Hal Smith for
valuable comments.


\begin{thebibliography}{99}                                                                                               %


\bibitem {An:As}G.~E.~Andrews and R.~A.~Askey, \textit{Classical orthogonal
polynomials\/}, in: \textsl{\textquotedblleft Polyn\^{o}mes orthogonaux et
applications\textquotedblright\/}, Lecture Notes in Math. \textbf{1171},
Springer-Verlag, 1985, pp.~36--62.

\bibitem {An:As:Ro}G.~E.~Andrews, R.~A.~Askey, and R.~Roy, \textsl{Special
Functions\/}, Cambridge University Press, Cambridge, 1999.

\bibitem {Arrighini:Durante}G.~P.~Arrighini and N.~L.~Durante, \emph{More on
the quantum propagator of a particle in a linear potential\/}, Am. J. Phys.
\textbf{64} (1996) \#~8, 1036--1041.

\bibitem {Askey}R.~A. Askey, \textsl{Orthogonal Polynomials and Special
Functions\/}, CBMS--NSF Regional Conferences Series in Applied Mathematics,
SIAM, Philadelphia, Pennsylvania, 1975.

\bibitem {As:Ra:Su}R. A. Askey, M. Rahman, and S. K. Suslov, \emph{On a
general }$q$\emph{-Fourier transformation with nonsymmetric kernels\/}, J.
Comp. Appl. Math. \textbf{68} (1996), 25--55.

\bibitem {Ba}W.~N.~Bailey, \textsl{Generalized Hypergeometric Series\/},
Cambridge University Press, Cambridge, 1935.

\bibitem {Beauregard}L.~A.~Beauregard, \emph{Propagators in nonrelativistic
quantum mechanics\/},~Am. J. Phys. \textbf{34} (1966), 324--332.

\bibitem {Bet:Cin:Kai:Cas}L.~M.~A.~Bettencourt, A.~Cintr\'{o}n-Arias,
D.~I.~Kaiser, and C.~Castillo-Ch\'{a}vez, \emph{The power of a good idea:
Quantitative modeling of the spread of ideas from epidemiological
models\/},~Phisica~A \textbf{364} (2006), 513--536.

\bibitem {Bo:Shi}N.~N.~Bogoliubov and D.~V.~Shirkov, \textsl{Introduction to
the Theory of Quantized Fields\/}, third edition, John Wiley \& Sons, New
York, Chichester, Brisbane, Toronto, 1980.

\bibitem {Brown:Zhang}L.~S.~Brown and Y.~Zhang, \emph{Path integral for the
motion of a particle in a linear potential\/}, Am. J. Phys. \textbf{62} (1994)
\# 9, 806--808.

\bibitem {Charlier}C.~V.~L.~Charlier, \emph{\"{U}ber die darstellung
willk\"{u}rlicher Funktionen\/}, Arkiv f\"{o}r Matematik, Astronomi och Fysik
\textbf{2} (1905-1906), No.~20, 35~pp.

\bibitem {Chihara}T.~S.~Chihara, \textsl{An Introduction to Orthogonal
Polynomials\/}, Gordon and Breach, New York, 1978.

\bibitem {Dav}A.~S.~Davydov, \textsl{Quantum Mechanics\/}, Pergamon Press,
Oxford and New York, 1965.

\bibitem {Erd}A.~Erd\'{e}lyi, \textsl{Higher Transcendental Functions\/},
Vol.~II, A.~Erd\'{e}lyi, ed., McGraw--Hill, 1953.

\bibitem {ErdInt}A.~Erd\'{e}lyi, \textsl{Tables of Integral Transforms\/},
Vols. I--II, A.~Erd\'{e}lyi, ed., McGraw--Hill, 1954.

\bibitem {FeynmanPhD}R.~P.~Feynman, \emph{The Principle of Least Action in
Quantum Mechanics\/}, Ph.~D. thesis, Princeton University, 1942; reprinted in:
\textsl{\textquotedblleft Feynman's Thesis -- A New Approach to Quantum
Theory\textquotedblright\/}, (L.~M.~Brown, Editor), World Scientific
Publishers, Singapore, 2005, pp.~1--69.

\bibitem {Feynman}R.~P.~Feynman, \emph{Space-time approach to non-relativistic
quantum mechanics\/}, Rev. Mod. Phys. \textbf{20} (1948)~\#~2, 367--387;
reprinted in: \textsl{\textquotedblleft Feynman's Thesis -- A New Approach to
Quantum Theory\textquotedblright\/}, (L.~M.~Brown, Editor), World Scientific
Publishers, Singapore, 2005, pp.~71--112.

\bibitem {Feynman49a}R.~P.~Feynman, \emph{The theory of positrons\/}, Phys.
Rev. \textbf{76} (1949)~\#~6, 749--759.

\bibitem {Feynman49b}R.~P.~Feynman, \emph{Space-time approach to quantum
electrodynamics\/}, Phys. Rev. \textbf{76} (1949)~\#~6, 769--789.

\bibitem {Fey:Hib}R.~P.~Feynman and A.~R.~Hibbs, \textsl{Quantum Mechanics and
Path Integrals\/}, McGraw--Hill, New York, 1965.

\bibitem {Flu}S.~Fl\"{u}gge, \textsl{Practical Quantum Mechanics\/},
Springer--Verlag, Berlin, 1999.

\bibitem {Gasper73}G.~Gasper, \emph{Nonnegativity of a discrete Poisson kernel
for the Hahn polynomials\/}, J. Math. Anal. Appl. \textbf{42} (1973), 438--451.

\bibitem {Gottf:T-MY}K.~Gottfried and T.-M.~Yan, \textsl{Quantum Mechanics:
Fundamentals\/}, second edition, Springer--Verlag, Berlin, New York, 2003.

\bibitem {Holstein97}B.\ R.\ Holstein, \emph{The linear potential
propagator\/}, Am. J. Phys. \textbf{65} (1997)~\#5, 414--418.

\bibitem {Holstein}B.\ R.\ Holstein, \emph{The harmonic oscillator
propagator\/}, Am. J. Phys. \textbf{67} (1998)~\#7, 583--589.

\bibitem {Howland}J.~Howland, \emph{Scattering theory for Hamiltonians
periodic in time\/}, Indiana Univ. Math. J. \textbf{28} (1979)~\# 3, 471--494.

\bibitem {Jafaev}D.~R.~Jafaev, \emph{Wave operators for the Schr\"{o}dinger
equation\/}, [in Russian] Teoret. Mat. Fiz. \textbf{45} (1980)~\#2, 224--234.

\bibitem {La:Lif}L.~D.~Landau and E.~M.~Lifshitz, \textsl{Quantum Mechanics:
Nonrelativistic Theory\/}, Pergamon Press, Oxford, 1977.

\bibitem {Maslov:Fedoriuk}V.~P.~Maslov and M.~V.~Fedoriuk,
\textsl{Semiclassical Approximation in Quantum Mechanics\/}, Reidel,
Dordrecht, Boston, 1981.

\bibitem {Me:Co:Su}M.~Meiler, R.~Cordero--Soto, and S.~K.~Suslov,
\emph{Solution of the Cauchy problem for a time-dependent Schr\"{o}dinger
equation\/}, arXiv: 0711.0559v4 [math-ph] 5 Dec 2007.

\bibitem {Meixner34}J.~Meixner, \emph{Orthogonale Polynomsysteme mit
einenbesonderen Gestalt der erzeugenden Funktion\/}, J. London Math. Soc.
\textbf{9} (1934), 6--13.

\bibitem {Meixner38}J.~Meixner, \emph{Erzeugende Funktionen der Charlierschen
Polynome\/}, Mathematische Zeitscrift \textbf{44} (1939) \#1, 531--535.

\bibitem {Meixner42}J.~Meixner, \emph{Unformung gewisser Reihen, deren Glieder
Produkte hypergeometrische Funktionen sind\/}, Deutsche Math. \textbf{6}
(1942), 341--349.

\bibitem {Merz}E.~Merzbacher, \textsl{Quantum Mechanics\/}, third edition,
John Wiley \& Sons, New York, 1998.

\bibitem {Mes}A.~Messia, \textsl{Quantum Mechanics\/}, two volumes, Dover
Publications, New York, 1999.

\bibitem {Naibo:Stef}V.~Naibo and A.~Stefanov, \emph{On some Schr\"{o}dinger
and wave equations with time dependent potentials\/}, Math. Ann. \textbf{334}
(2006) \# 2, 325--338.

\bibitem {Nardone}P.~Nardone, \emph{Heisenberg picture in quantum mechanics
and linear evolutionary systems\/}, Am. J. Phys. \textbf{61} (1993) \# 3, 232--237.

\bibitem {Ni:Su:Uv}A.~F.~Nikiforov, S.~K.~Suslov, and V.~B.~Uvarov,
\textsl{Classical Orthogonal Polynomials of a Discrete Variable\/},
Springer--Verlag, Berlin, New York, 1991.

\bibitem {Ni:Uv}A.~F.~Nikiforov and V.~B.~Uvarov, \textsl{Special Functions of
Mathematical Physics\/}, Birkh\"{a}user, Basel, Boston, 1988.

\bibitem {Palio:Mead}J.~D.~Paliouras and D.~S.~Meadows, \textsl{Complex
Variables for Scientists and Engineers\/}, second edition, Macmillan
Publishing Company, New York and London, 1990.

\bibitem {Rainville}E.~D.~Rainville, \textsl{Special Functions\/}, Chelsea
Publishing Company, New York, 1960.

\bibitem {Ra:Su}M.~Rahman and S.~K.~Suslov, \emph{Singular analogue of the
Fourier transformation for the Askey--Wilson polynomials}\textit{\/}, in:
\textsl{Symmetries and Integrability of Difference Equations}, D.~Levi,
L.~Vinet, and P.~Winternitz, eds., CRM Proceedings \& Lecture Notes, Vol.~9,
Amer. Math. Soc., 1996, pp.~289--302.

\bibitem {Robinett}R.~W.~Robinett, \emph{Quantum mechanical time-development
operator for the uniformly accelerated particle\/}, Am. J. Phys. \textbf{64}
(1996) \#6, 803--808.

\bibitem {Rod:Schlag}I.~Rodnianski and W.~Schlag, \emph{Time decay for
solutions of Schr\"{o}dinger equations with rough and time-dependent
potentials}\textit{\/}, Invent. Math. \textbf{155} (2004)~\# 3, 451--513.

\bibitem {Schiff}L.~I.~Schiff, \textsl{Quantum Mechanics\/}, third edition,
McGraw-Hill, New York, 1968.

\bibitem {Schlag}W.~Schlag, \emph{Dispersive estimates for Schr\"{o}dinger
operators: a survay}\textit{\/}, arXiv: math/0501037v3 [math.AP] 10 Jan 2005.

\bibitem {SteinHarm}E.~M.~Stein, \textsl{Harmonic Analysis: Real-Variable
Methods, Orthogonality, and Oscillatory Integrals\/}, Princeton University
Press, Princeton, New Jersey, 1993.

\bibitem {SusAlg}S.~K.~Suslov, \emph{An algebra of integral operators}%
\textit{\/}, Electronic Transactions on Numerical Analysis (ETNA), \textbf{27}
(2007), 140--155.

\bibitem {Sze}G.~Szeg\H{o}, \textsl{Orthogonal Polynomials\/}, Amer. Math.
Soc. Colloq. Publ., Vol. 23, Rhode Island, 1939.

\bibitem {Thomber:Taylor}N.~S.~Thomber and E.~F.~Taylor, \emph{Propagator for
the simple harmonic oscillator\/},~Am. J. Phys. \textbf{66} (1998) \# 11, 1022--1024.

\bibitem {Vil}N.~Ya.~Vilenkin, \textsl{Special Functions and the Theory of
Group Representations\/}, American Mathematical Society, Providence, 1968.

\bibitem {Wi}N. Wiener, \textsl{The Fourier Integral and Certain of Its
Applications\/}, Cambridge University Press, Cambridge, 1933; Dover edition
published in 1948.

\bibitem {Yajima}K.~Yajima, \emph{Scattering theory for Schr\"{o}dinger
equations with potentials periodic in time}\textit{\/}, J.~Math. Soc. Japan
\textbf{29} (1977) \# 4, 729--743.
\end{thebibliography}
\end{document}